\documentclass[10pt,twoside]{article}
\usepackage{own_sngl}
\usepackage[dvips]{graphics}
\newcommand{\oversim}[2]{\protect{\mbox{\lower0.5ex\vbox{%
   \baselineskip=0pt\lineskip=0.2ex
   \ialign{$\mathsurround=0pt #1\hfil##\hfil$\crcr#2\crcr\sim\crcr}}}}} 
\newcommand{\simgreat}{\mbox{$\,\mathrel{\mathpalette\oversim>}\,$}} 
\newcommand{\simless} {\mbox{$\,\mathrel{\mathpalette\oversim<}\,$}} 
\slugcomment{{\em {\bf MNRAS}, submitted 04.05.00; accepted 22.09.00}}
\lefthead{Kroupa, Aarseth \& Hurley}
\righthead{Star Cluster Formation}
\begin{document}
%
\title {The Formation of a Bound Star Cluster:\\
From the Orion Nebula Cluster to the Pleiades}

\author {Pavel Kroupa$^1$, Sverre Aarseth$^2$ and Jarrod Hurley$^2$\\
\medskip
\small{$^1$Institut f{\"u}r Theoretische Physik und Astrophysik, 
Universit{\"a}t Kiel\\ Olshausenstr.~40, D-24118 Kiel, Germany\\
$^2$ Institute of Astronomy, University of Cambridge\\ Madingley Road,
Cambridge CB3~OHA, England
%
}}
\begin{abstract}
\noindent 
Direct $N$-body calculations are presented of the formation of
Galactic clusters using {\sc GasEx}, which is a variant of the code
{\sc Nbody6}.  The calculations focus on the possible evolution of the
Orion Nebula Cluster (ONC) by assuming that the embedded OB stars
explosively drove out $2/3$ of its mass in the form of gas about
0.4~Myr ago. A bound cluster forms readily and survives for 150~Myr
despite additional mass loss from the large number of massive stars,
and the Galactic tidal field. This is the very first time that cluster
formation is obtained under such realistic conditions.  The cluster
contains about $1/3$ of the initial $10^4$ stars, and resembles the
Pleiades Cluster to a remarkable degree, implying that an ONC-like
cluster may have been a precursor of the Pleiades. This scenario
predicts the present expansion velocity of the ONC, which will be
measurable by upcoming astrometric space missions.  These missions
should also detect the original Pleiades members as an associated
expanding young Galactic-field sub-population.  The results arrived at
here suggest that Galactic clusters form as the nuclei of expanding OB
associations.

The results have wide implications, also for the formation of globular
clusters and the Galactic field and halo stellar populations. In view
of this, the distribution of binary orbital periods and the mass
function within and outside the model ONC and Pleiades is quantified,
finding consistency with observational constraints. Advanced mass
segregation is evident in one of the ONC models.  The calculations
show that the primordial binary population of both clusters could have
been much the same as is observed in the Taurus--Auriga star forming
region. The computations also demonstrate that the binary proportion
of brown dwarfs is depleted significantly for all periods, whereas
massive stars attain a high binary fraction.
\end{abstract}


\section{INTRODUCTION} 
\label{sec:intro}
\noindent
The Orion Nebula Cluster (ONC) is a rather peculiar object. It is
extremely young ($< 2.5$~Myr; Hillenbrand 1997; Palla \& Stahler
1999), contains 4000--10000 stars within a region with a diameter of
at least 5~pc and has a very high central number density of about
$10^{4.7}$~stars/pc$^3$ (McCaughrean \& Stauffer 1994). It also has a
velocity dispersion that is too large for it's known mass (Jones \&
Walker 1988).  Thus, it ought to be expanding, which may be a result
of very recent ($\simless 0.5$~Myr ago) loss of a substantial amount
of gas when the central OB stars ``ignited''. It has been debated for
a long time whether the ONC actually is a proto-Galactic cluster, or
if it is in the process of forming an unbound OB association
(e.g. Zinnecker, McCaughrean \& Wilking 1993). Indeed, a large
fraction of the ONC stars are optically visible and thus not embedded
in gas (Hillenbrand \& Hartmann 1998), and its central region, the
Trapezium Cluster, contains less than $10\,M_\odot$ in gas (Wilson et
al. 1997).  The large velocity dispersion implies that an unbound
association is likely to result, but a recent investigation of the set
of possible dynamical solutions strengthens the possibility that the
ONC is close to forming a bound cluster (Kroupa 2000a). There is no
discernible sub-structure in the Trapezium so that it is at
least a few crossing times ($t_{\rm cross}$) old
(Bate, Clarke \& McCaughrean 1998).

Under which conditions some embedded clusters survive to become bound
entities, such as the Pleiades Cluster, remains something of a mystery
(see Elmegreen et al. 2000; Clarke, Bonnell \& Hillenbrand 2000 for
reviews).  It is known though that in the process of forming an
embedded cluster a relatively large mass of gas, typically $\simgreat
500\,M_\odot$, is squeezed into a volume of typically one~pc$^3$. The
gas dynamics is complex, but once stars form, they remove a
substantial fraction of this gas through outflows and
winds. Especially in those circumstances when~O stars form and start
shining, gas removal can be expected to be very rapid and faster than
the dynamical crossing time of the embedded cluster (Whitworth 1979).
Observations indicate that embedded massive stars have dynamically
young ($\simless10^4$~yr) outflows that contain more mass than the
powering star (Churchwell 1997, 1999). In addition to the expansion of
the HII region by virtue of it's overpressure, and the mass loss from
the central region containing the massive star through the outflow, a
massive star produces a powerful wind that is very effective in
expelling mass from the cluster (Garay \& Lizano 1999).

Star-formation is typically less than 40~per cent efficient (Lada
1999; Elmegreen et al. 2000; Clarke et al. 2000), and once more than
50~\% of the mass of a cluster is removed instantaneously, an unbound
association is deemed to result, unless the cluster was {\it beginning
to} collapse, or was virialised after collapse, at the instant of gas
removal.  Such results that laid the conceptual framework for this
work and which were obtained at a time when infrared observations did
not exist to constrain the star-formation efficiency (sfe) or embedded
cluster morphology and membership, were arrived at using either a
direct $N$-body code being limited to $\le100$ stars (Lada, Margulis
\& Dearborn 1984), or necessarily simplifying analytical arguments
(Hills 1980; Elmegreen 1983; Mathieu 1983) with more recent
generalisations (Pinto 1987; Verschueren \& David 1989; Verschueren
1990). A particularly interesting result that emerged from this early
ground-breaking work was the realization that, unless the sfe is
larger than~50~per cent, Galactic (i.e. 'open') clusters can not form
containing OB stars, because these drive out the gas faster than the
dynamical crossing time leading to complete dispersion of the stellar
system.  With a state-of-the-art $N$-body code and advanced
observational constraints, it is now possible to re-visit this problem
with the aim of investigating the likely fate of the ONC as a first
application.

The ONC can be in one of three dynamical states (Kroupa 2000a): (i) in
virial equilibrium, (ii) collapsing, or just after collapse and in or
past the associated violent relaxation phase, or (iii) expanding.
Case (i) certainly leads to a bound cluster, but is very unlikely
because it implies an sfe of essentially 100~\% ($\epsilon=1$), which
has never been observed. Case (ii) may occur if stars form out of the
molecular cloud with a smaller-than-virial velocity-dispersion, so
that the stellar system contracts. Rapid gas removal will then leave a
bound cluster, {\it if} the gas is removed rapidly at onset of
collapse, or after virialisation.  If even a relatively small fraction
of the gas is removed during the main collapse phase, however, then an
unbound association is likely to result (Hills 1980). In this case the
observer should be seeing the ONC in the special time just before or
after maximum central density with a radial infall or expanding bulk
motion. However, such a scenario requires star formation to be
finished within less than one proto-cluster crossing time ($\simless
0.4$~Myr for the ONC), and thus a physically implausible degree of
synchronisation across the whole cluster-forming cloud.

In reality, stars appear to continue forming in a cluster for a
few~Myr until the gas is expelled (Palla \& Stahler 2000), and once a
star forms it decouples from the gas and either falls towards the
potential minimum or is already on a ``virial equilibrium orbit''. In
any case, the accumulating stellar system will be virialising
constantly as new stars are added, while the stars formed less than a
crossing time ago may be falling towards the cluster centre. Gas-drag
may cause the stellar system to contract enhancing the effective sfe
(Saiyadpour, Deiss \& Kegel 1997). Thus, most of the cluster is likely
to be in virial equilibrium when the first massive stars ``ignite''
with devastating effects for the residual (but still mass-dominating)
gas.  This scenario is supported by various observations which show
star-formation on-going in a clustered environment (e.g. Megeath et
al. 1996).

In this paper the one scenario (iii) that is {\it least likely to form
a bound cluster} and which is the {\it most intuitive}, is studied in
the realistic situation of very rapid gas removal using a newly
completed extension of the code {\sc Nbody6} (Aarseth 1999), {\sc
GasEx}.  It is a modified version of {\sc Nbody6}, by incorporating
the gas in an embedded cluster as a time-evolving background
potential. The {\it essential physics} of gas expulsion is thus
described in a computationally-efficient way, while retaining the
necessary highly accurate treatment of point-mass stellar encounters.
{\sc GasEx} also allows the setup of a realistic primordial binary
population according to Kroupa (1995b), and a new five-part power-law
IMF-generating routine extends the mass range into the brown dwarf
(BD) regime.

Calculations are performed of clusters that initially resemble the ONC
and which contain twice as much mass in gas than in stars. After
0.6~Myr the gas is removed on a time-scale given by the velocity of
the heated ionised gas (10~km/s, e.g. Hills 1980).  Despite the
essentially instantaneous gas loss, roughly $1/3$ of the cluster
remains bound and forms a cluster remarkably similar to the Pleiades.
This result differs from previous work, as is discussed in
Section~\ref{sec:disc}.

The code is described in Section~\ref{sec:code}, and the adopted
initial conditions are detailed in
Section~\ref{sec:ic}. Section~\ref{sec:res} contains the results and
predictions. A discussion is provided in Section~\ref{sec:disc}, and
the paper is concluded with Section~\ref{sec:concl}.

\section{THE CODE}
\label{sec:code}
\noindent
A few points concerning {\sc Nbody6} are stressed, and the inclusion
of the gas potential is described in this section. 

\subsection{The code {\sc Nbody6}}
\noindent
The stellar-dynamical interactions are treated with {\sc Nbody6},
which is described elsewhere in detail (Aarseth 1999, 2000).

A direct $N$-body code must deal efficiently with a range of dynamical
time-scales spanning many orders of magnitude. To achieve this,
special mathematical techniques are employed to transform the
space-time coordinates of closely interacting stars such
that the resulting equations of motion of the sub-system are regular
(Mikkola \& Aarseth 1993), that is, the calculation can proceed
accurately through the near time-step singularity that occurs during an
encounter.  This {\it regularisation} extends from the two-body case
to a chain containing up to six members. Perturbations acting on the chain
from close non-chain-members are incorporated, and allowance is made
for approaching stars or binaries to become new members, and old
members to be ejected from the usually unstable small-body group.

In addition to the efficient treatment of close encounters and
transient compact sub-systems, the code assigns a neighbour radius
for each star which typically contains the nearest $\sqrt{N}$ stars,
where $N$ is the total number
of stars in the cluster. The {\it irregular} forces from these
neighbours are computed on the smallest time-step of their central star,
and the {\it regular} forces from the stars lying outside the
neighbour sphere are added at less frequent intervals.  The {\it
Ahmad-Cohen} (Ahmad \& Cohen 1973) neighbour-scheme makes the CPU time
scale approximately with $N^{1.6}$ rather than the usual $N^2$ in a
``primitive'' direct $N$-body code.  Each star has it's own regular
and irregular time-step, which is adjusted throughout the computation,
but is alway commensurate with 2. That is, {\it block-time-steps} are
used in this version, as opposed to the basic individual time-step
scheme in {\sc Nbody6}'s precursor, {\sc Nbody5}, which was
previously applied to the problem of young clusters (e.g. Kroupa, Petr
\& McCaughrean 1999, hereinafter KPM).  Block-time-steps are used in
connection with Hermite integration, and
CPU-timing tests show that it is more accurate and at least as fast as
{\sc Nbody5} for the same number of integration steps.

State-of-the-art stellar evolution is incorporated using analytical
fitting functions (Hurley, Pols \& Tout 2000).  These functions allow
the generation of synthetic HR diagrams for a wide range of the
metallicity.  However, in these simulations we adopt the standard
solar value (Z = 0.02).  The resulting mass loss is implemented in a
nearly continuous manner which facilitates an energy-conserving scheme
(Aarseth 1999).  Although the code contains a simple scheme for tidal
circularisation, this process has not been included up to now, with
consequent neglect of Roche-lobe mass transfer.  The code assigns kick
velocities when the massive stars explode, so that high-velocity
neutron stars are readily produced. This, however, is not a topic of
the present study.

\subsection{The gas}
\label{sec:gas}
\noindent
To represent the essential physics of the dynamical evolution of an
embedded cluster when the gas is removed, a time-varying analytical
spherical background potential is added, following the pioneering
study of Lada et al. (1984). This avoids the daunting task of a
cluster-scale three-dimensional magneto-hydrodynamical computation
with stellar feedback, but includes the important dynamical effect of
reducing the containing potential as the gas is blown out. That the
analytical approximation for the gas leads to physically realistic
behaviour is concluded by Geyer \& Burkert (2000), who apply
collisionless $N$-body computations to this problem assuming an
analytical gas as well as a more realistic treatment with
smoothed-particle hydrodynamics (SPH).

Both the embedded star cluster and the parent cloud are
assumed to be Plummer spheres. A star with a position vector ${\bf R}$
relative to the centre of the cluster experiences an acceleration from
the background potential,
\begin{equation}
{\bf a} = -{G\,M_{\rm g}(t) \over (R^2 + R_{\rm
pl,g}(t)^2)^{3/2} }\,{{\bf R}\over R},
\end{equation}
where $M_{\rm g}(t)$ and $R_{\rm pl,g}(t)$ are the time-varying mass
and Plummer radius of the gas cloud, and G is the gravitational
constant. The background potential is assumed to begin evolving after
a delay time $t_{\rm D}$ according to
\begin{equation}
M_{\rm g}(t) = M_{\rm g}(0)\,
         e^{-\left({(t-t_{\rm D})\over\tau_{\rm M}}\right)},
         t\ge t_{\rm D},
\label{eq:Mt}
\end{equation}
and similarly the radius evolves as
\begin{equation}
R_{\rm pl,g}(t) = R_{\rm pl,g}(0)\,
       \left[1+\left({(t-t_{\rm D})\over\tau_{\rm
       R}}\right)^{1/2}\right], t\ge t_{\rm D}, 
\end{equation}
where $\tau_{\rm M}$ and $\tau_{\rm R}$ are the time-scales for mass
and radius evolution, respectively.  In the present application
$R_{\rm pl,g}(t)$ remains unchanged.

In general, the changes to $M_{\rm g}$ and $R_{\rm pl,g}$ are applied
at constant time intervals, $\delta t_{\rm a}<<{\rm min}(\tau_{\rm
M},\tau_{\rm R})$.  To conserve energy at every adjustment the
resulting change in potential energy of the system of $N$ stars is
obtained by summing over the changes of each star with mass $m_{\rm
i}$, giving $\delta \Phi = \sum_{i=1}^{N}m_{\rm i}\,\delta\Phi_{\rm i}$,
where
\begin{equation}
\delta\Phi_{\rm i} = -{G\,M_{\rm g}(t+\delta t_{\rm a}) \over 
         \left(R_{\rm i}^2 + R_{\rm pl,g}^2(t+\delta t_{\rm a})\right)^{1/2}}
             +
                      {G\,M_{\rm g}(t) \over 
         \left(R_{\rm i}^2 + R_{\rm pl,g}^2(t)\right)^{1/2}}.
\end{equation}
The contributions by the gas potential are neglected once $M_{\rm
g}(t)<10^{-5}\,M_\odot$.

\section{INITIAL CONDITIONS}
\label{sec:ic}
\noindent
The initial properties of the stellar population and of the embedded
cluster models are described. 

\subsection{The initial stellar and binary population}
\label{sec:initpop}
\noindent
Initial stellar masses are distributed according to the universal
three-part power-law IMF (Kroupa 2000c) with lower and upper mass
limits $m_{\rm l}=0.01\,M_\odot$ and $m_{\rm u}=50\,M_\odot$,
respectively:
\begin{equation}
\xi(m) \propto m^{-\alpha_i},
\end{equation}
where
\begin{equation}
          \begin{array}{l@{\quad\quad,\quad}l}
\alpha_0 = +0.3   &0.01 \le m/M_\odot < 0.08, \\
\alpha_1 = +1.3   &0.08 \le m/M_\odot < 0.50, \\
\alpha_2 = +2.3   &0.50 \le m/M_\odot. \\
          \end{array}
\label{eq:imf}
\end{equation}
Here, $\xi(m)\,dm$ is the number of stars in the mass interval $m$ to
$m+dm$.  An often useful description is achieved via the {\it
logarithmic} IMF,
\begin{equation}
\xi_{\rm L}(m)=\xi(m)\,m\,{\rm ln}10,
\label{eq:imfl}
\end{equation}
where $\xi_{\rm L}$ is the number of stars in the logarithmic mass
interval log$_{10}m$ to log$_{10}m+dm$. This is the form used in later
figures of this paper.

The above IMF leads to the following stellar population: 37~\% BDs
($0.01-0.08\,M_\odot$) contributing 4.3~\% to the stellar mass, 48~\%
M~dwarfs ($0.08-0.5\,M_\odot$) contributing 28~\% mass, 8.9~\%
``K''~dwarfs ($0.5-1.0\,M_\odot$) contributing 17~\% mass, 5.7~\%
``intermediate mass (IM) stars'' ($1.0-8.0\,M_\odot$) contributing
34~\% mass, and 0.37~\% ``O'' stars ($>8\,M_\odot$) contributing 17~\%
mass.

The total binary proportion is 
\begin{equation}
f_{\rm tot} = {N_{\rm bin}\over(N_{\rm bin}+N_{\rm sing})},
\label{eq:f}
\end{equation}
where $N_{\rm bin}$ and $N_{\rm sing}$ are the number of binary and
single-star systems, respectively.  

Initially, all stars are assumed to be in binary systems ($f_{\rm
tot}=1$) with component masses $m_1, m_2$ chosen randomly from the
IMF. This gives approximately a flat distribution of mass-ratios,
$q=m_2/m_1 \le1$, with a maximum at $q=1$ as a result of the model
adopted by Kroupa (1995b) for system-internal processes, such as
star-disk interactions, tidal circularisation and mass ejection that
redistribute mass, angular momentum and energy within the binary system
while it is still very young ($<10^5$~yr). For convenience, this
evolution is collectively referred to as 'pre-main sequence
eigenevolution', to differentiate it from the externally-induced
perturbations from nearby stars.  Random pairing over the entire mass
range implies that most massive primaries will initially have M~dwarf
and BD~companions, which may not be consistent with observational
constraints. Their numbers are so small though, that they do not
significantly alter the overall mass-ratio distribution, $f_q$. The
fast sinking to the cluster centre leads to rapid (within a few
$t_{\rm cross}$) changes in $f_q$ for massive primaries as they pair
up with massive secondaries near the cluster core, but details await a
further study.

The birth period distribution, with periods ranging from about 1~d to
$10^9$~d, is constructed following Kroupa (1995b, eq.8), and pre-main
sequence eigenevolution gives the observed correlation between orbital
period and eccentricity for short-period systems with stellar masses
in the range $0.08-1.1\,M_\odot$.  Only systems with a period
$P\simless10^4$~d are affected, the overall distribution of masses not
being changed greatly. As a result, some binaries merge, so that at
$t=0$ the binary proportion is slightly below unity.  The {\it initial
binary proportion} is reduced further through crowding, that is,
immediate disruption because some long-period systems have companion
separations larger than the system--system distances in the initial
cluster, and because of the cluster tidal field, giving the true {\it
initial binary proportion}.

\subsection{Cluster models}
\label{sec:clmods}
\noindent
The star clusters are assumed to have spherical Plummer
number-density profiles initially (e.g. Aarseth, H\'enon \& Wielen 1974) with
half-mass radii $R_{0.5}$ and mass $M_{\rm st}$.  The true initial
conditions of a cluster are not known, and are likely to be very
complex with significant sub-structure, which evolves on a crossing
time-scale with star formation continuing for some time until the gas
is expelled (e.g. Klessen, Burkert \& Bate 1998).  A more detailed
model will not be attempted until the evolution of the simpler spherical
case with gas expulsion as set up here is understood.

The number of stars, $N_{\rm in}$, and initial central density,
$\rho_{\rm C}$, are taken from Kroupa (2000a), where possible initial
configurations of the ONC are constrained.  $N_{\rm in}=10^4$ is an
upper limit to the initial number of stars in the ONC, and in expanding
models $\rho_{\rm C}$ must be at least as large as the current ONC
value.

Stellar masses and binary properties are assumed to be uncorrelated
with their location in the cluster, giving an average stellar and
system mass that is constant with $R$. Again, this assumption is not
likely to be realistic because observational evidence and theoretical
arguments suggest that massive stars may form near cluster centres
(Bonnell, Bate \& Zinnecker 1998). The present assumption is made
because it allows measurement of the rapidity with which dynamical
mass segregation occurs when calculated with a high-precision $N$-body
code (touched upon in Fig.~\ref{fig:mfn09c}). This, however, will be
the focus of another contribution.

The gas is assumed to have the same spatial distribution as the stars
with a half-mass radius $R_{0.5,{\rm g}}=R_{0.5}$. It is ``expelled''
by reducing its mass, $M_{\rm g}(t)$, at time $t\ge t_{\rm D}$ on a
time-scale $\tau_{\rm M}$.  $M_{\rm g}(0)=2\,M_{\rm st}$, i.e. an
sfe $\epsilon\approx0.3$, is assumed.

The Galactic tidal field is instrumental in tidally truncating an
expanding cluster.  A standard Galactic tidal field is adopted using
the linearised approximation (Terlevich 1987) with Oort's constants
$A=14.4$~km/s/kpc and $B=-12.0$~km/s/kpc, and a mass density in the
solar neighbourhood of $0.11\,M_\odot$/pc$^3$. Stars are removed from
the $N$-body calculation once they reach a distance of 100~pc from
their cluster, but are kept in memory without advancing them on their
Galactic orbit, but with stellar evolution included, to facilitate data
reduction.

Two models are discussed here. The model parameters are listed in
Table~\ref{tab:mods}.

\section{RESULTS}
\label{sec:res}
\noindent 
The evolution of the two models is discussed, with a comparison to
some observational constraints available for both the ONC and the
Pleiades. These constraints include the structural and kinematic
properties, as well as the binary proportion and period
distributions. The section ends with a discussion of the stellar and
system mass functions.

Throughout this paper, $R$ refers to the 3D radial distance from the
density centre of the cluster, whereas $r$ is the projected 2D
distance. The projection onto the ``observational plane'' is
arbitrary, except that the vertical direction is always taken to be
perpendicular to the Galactic disc. Pictures of the distribution of
stars in the model clusters at $t=100$~Myr thus show flattened
clusters with shortest extension along the y-direction (c.f. Terlevich
1987; Portegies Zwart et al. 2000).

\subsection{Cluster properties vs observations}
\label{sec:res:bulk}
\noindent
Not much is known about the stellar-dynamical evolution of a cluster
for which the velocity dispersion is increased as a result of an
additional containing potential. In the cases considered here, the 3D
velocity dispersion $\sigma_{\rm 3D}=\sqrt{3}\,\sigma_{\rm 3D, st}$,
where $\sigma_{\rm 3D, st}$ is the velocity dispersion if no gas were
present. The crossing time is thus shortened by the factor $\sqrt{3}$,
and stellar interactions occur $\sqrt{3}$ times as often, because the
stellar density remains unchanged. This topic
requires more study, and one result of the present calculations
is that for example $f_{\rm tot}$ (eq.~\ref{eq:f}) decays in {\it the
same manner} as for a cluster without gas but the same density
distribution, which is somewhat surprising. Hence the increased
number of encounters is compensated by the reduced encounter duration.

Evaluating the mean radial distance of stars in various mass bins as a
function of time shows that mass segregation is very weak in model~A
at the time of gas expulsion, whereas it develops to be quite
pronounced in model~B, leading to a slight expansion of stars with
mass $m<8\,M_\odot$, as a result of the associated heating of the
cluster.  This will be re-visited briefly in Section~\ref{sec:res:mf}
below.

\begin{figure}
\plotfiddle{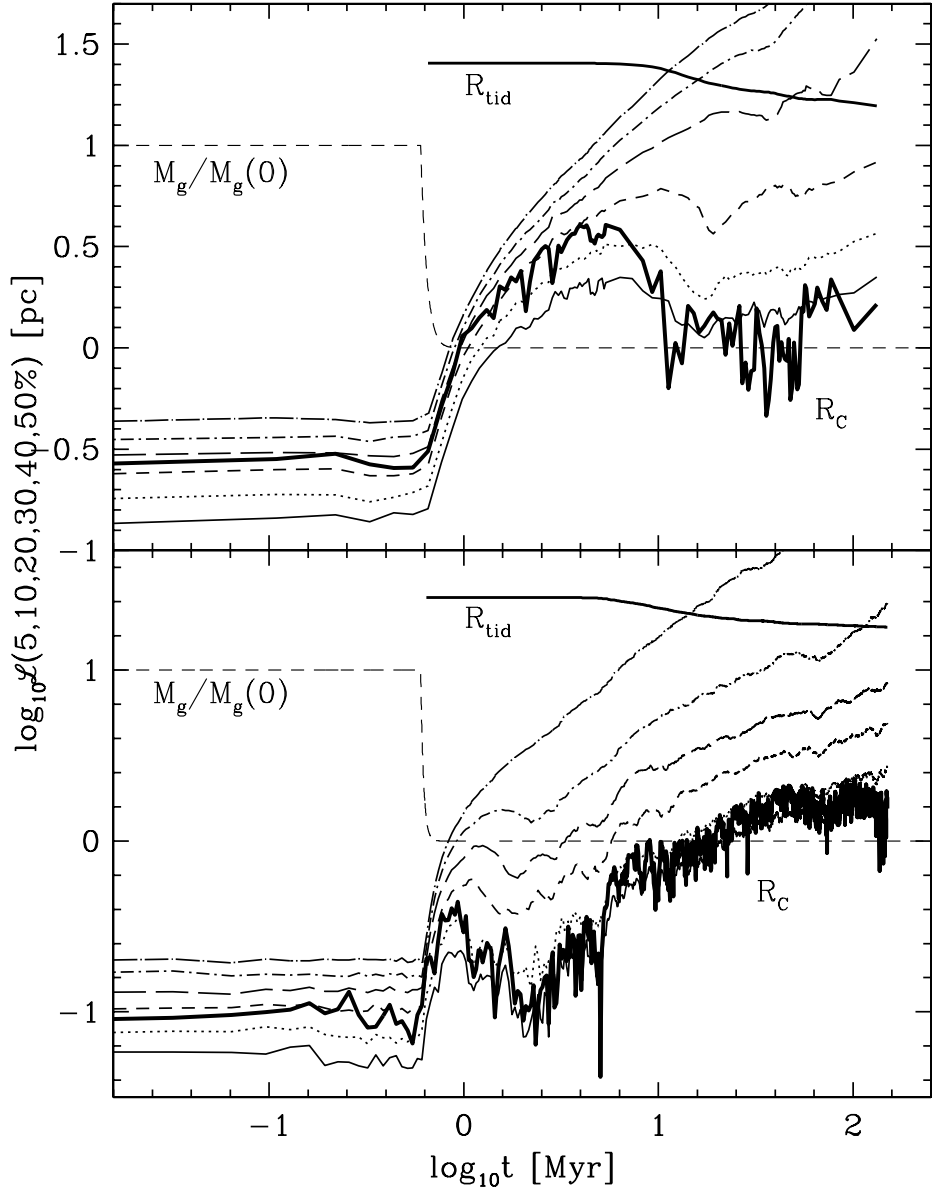}{15cm}{0}{64}{64}{-190}{-20}
\caption{The evolution of the 5,~10,~20...~50~per~cent Lagrange radii,
of the core radius (log$_{10}R_{\rm C}$: thick lower curve) and of the
approximate tidal radius (log$_{10}R_{\rm tid}$: thick upper curve).
The evolution of the gas mass (not log$_{10}$) is shown as the thin
dashed line, $M_{\rm g}(0)=7492\,M_\odot$ (model A) and
$8340\,M_\odot$ (model B). Upper panel is for model~A and the lower
panel for model~B (Table~\ref{tab:mods}).
\label{fig:lagr}}
\end{figure}

At time $t\ge t_{\rm D}=0.6$~Myr the gas mass evolves according to
eq.~\ref{eq:Mt} with $\tau_{\rm M}$ given in Table~\ref{tab:mods}.  In
both models, $M_{\rm g}(t)\approx0$ for $t\simgreat1$~Myr, the
background potential being insignificant thereafter, and most of the
original cluster expands as an OB association. Fig.~\ref{fig:lagr}
depicts this expansion, but also shows that between~20 and~30~\% of
the stellar population remains within the approximate tidal radius
(Binney \& Tremaine 1987),
\begin{equation}
R_{\rm tid}(t) = \left({M_{\rm st}(t)\over3\,M_{\rm
gal}}\right)^{1\over3}\,R_{\rm GC},
\label{eq:rtid}
\end{equation}
with $M_{\rm gal}=5\times10^{10}\,M_\odot$ being approximately the
Galactic mass enclosed within the distance of the Sun to the Galactic
centre, $R_{\rm GC}=8.5$~kpc. To estimate $R_{\rm tid}(t)$, $M_{\rm
st}(t)$ is calculated by summing only those stars which have $R(t)\le
2\,R_{\rm tid}(t-\delta t_{\rm op})$, where the data output time
interval $\delta t_{\rm op} << t_{\rm relax}(t)$.  $R_{\rm tid}(t)$ is
shown in Fig.~\ref{fig:lagr} for $t>0.6$~Myr merely as an aid to the
eye to facilitate a rough estimate of how many stars form the bound
cluster by 100~Myr ($R_{\rm tid}$ is not used to remove stars from the
$N$-body calculation). The bound cluster that forms from the expanding
OB association contracts with time as $R_{\rm tid}$ decreases, and the
density profiles at~100~Myr are discussed further below
(Fig.~\ref{fig:radpr_pl}).

The core radius, $R_{\rm C}$, is a useful quantity because it can be
constrained observationally much more easily than the half-mass radius
or tidal radius. In the $N$-body computations it is approximated by
calculating the density-weighted radius $R_{\rm HA}$ (Heggie \&
Aarseth 1992),
\begin{equation}
R_{\rm HA}^2 = {\sum_{i=1}^{N_{20}} R_{\rm i}^2\,\rho_{\rm i}^2 \over
               \sum_{\rm i=1}^{N_{20}} \rho_{\rm i}^2},
\end{equation}
where $\rho_{\rm i}= 3\,m_{\rm i,5}/(4\pi\,d_{\rm i,5}^3)$ is the
density around star~i estimated within the closest distance $d_{\rm
i,5}$ from star~i containing $n=5$ additional stars with combined mass
$m_{\rm i,5}$, and $R_{\rm i}$ is the distance of star~i from the
density centre of the cluster. The summation extends only over the
innermost 20~\% of all stars in the cluster, $N_{20}$, since this is
sufficient to ensure convergence. For a
comparison with the core radius, $R_{\rm C}$, obtained from fitting
King density profiles, as is often done for observed clusters such as
the ONC and the Pleiades,
\begin{equation}
R_{\rm C}\approx R_{\rm HA}/0.8,
\label{eq:rc}
\end{equation}
(Heggie \& Aarseth 1992; Giersz \& Spurzem 2000). The evolution of
$R_{\rm C} = R_{\rm HA}/0.8$ is shown in Fig.~\ref{fig:lagr}. It follows
the~5 and~10\% Lagrange radii rather well.

The formation of a bound cluster despite $\epsilon\approx0.3$,
despite explosive mass loss and additional mass-loss from evolving
massive stars, and despite the Galactic tidal field, is evident by a
contraction of $R_{\rm C}$ for $t\simgreat5$~Myr (model~A) and
$t\simgreat1$~Myr (model~B). Evolving massive stars cause $R_{\rm C}$
to expand in model~B for $t\simgreat 2.0$~Myr, and in both models
$R_{\rm C}\approx1.5$~pc when $t=150$~Myr is reached. The core radius
fluctuates by about 0.5~pc as a result of Poisson noise.  It is rather
remarkable that $R_{\rm C}$ ends up being so similar for both models
despite the significantly different initial density.

Thus, in the presence of a tidal field, models of the same cluster
mass converge to a similar structural state independent of the initial
central density used.  That is, after some time the inner regions
begin to feel the effect of the contracting tidal radius and
``self-similar'' evolution sets in: the core- and half-mass radii then
depend only on the total cluster mass even though, for each individual
model, the evolution path of the core radius to this point may be
quite different. This was evident in the cluster models with low-mass
stars computed by Kroupa (1995c), and in more realistic models of M67
by Hurley (2000) and Hurley et al. (2000b).
   
\begin{figure}
\plotfiddle{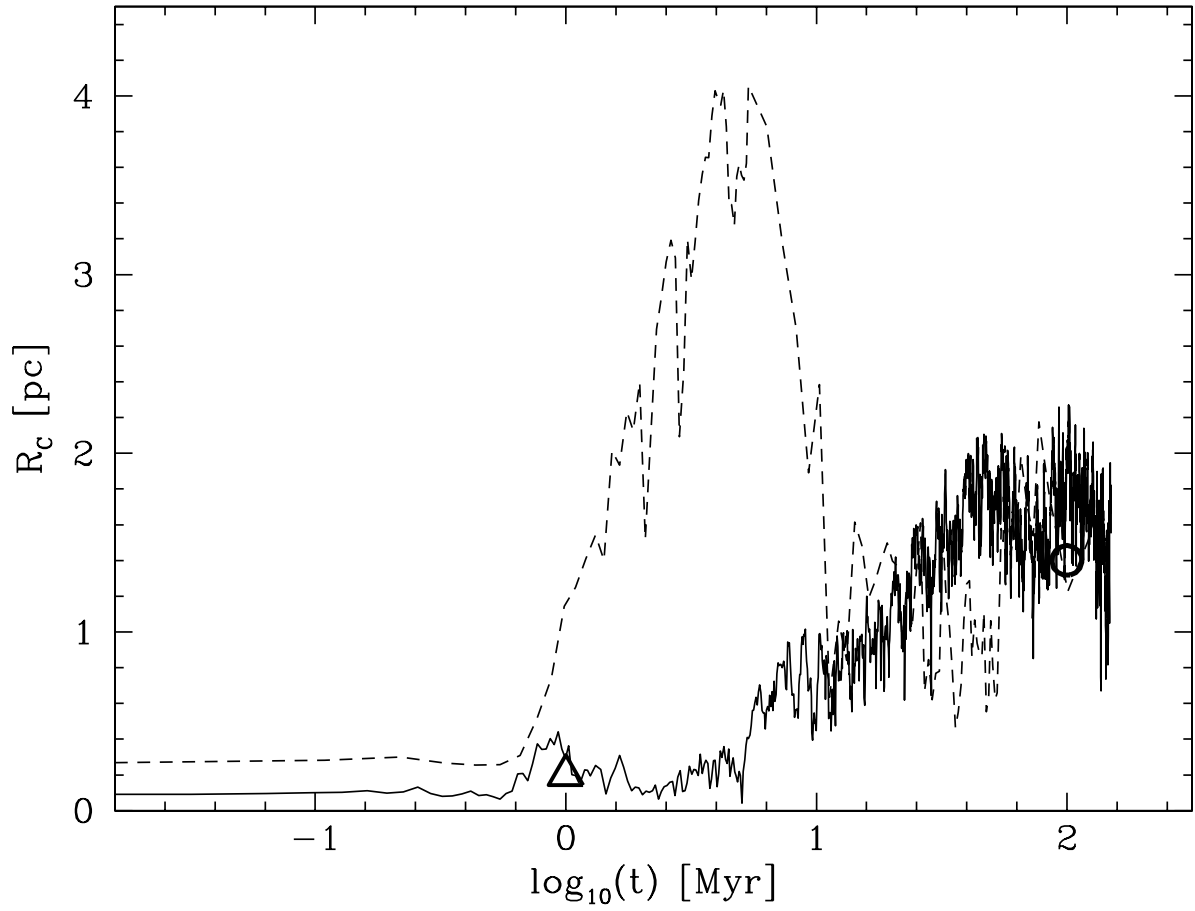}{15cm}{-90}{63}{63}{-250}{400}
\caption{The evolution of the core radius (eq.~\ref{eq:rc}, dashed
curve: model~A, solid curve: model~B).  The open circle is the
Pleiades datum (Raboud \& Mermilliod 1998), and the open triangle is
the core radius of the Orion Nebula Cluster (Hillenbrand \& Hartmann
1998).
\label{fig:rc}}
\end{figure}

\begin{figure}
\plotfiddle{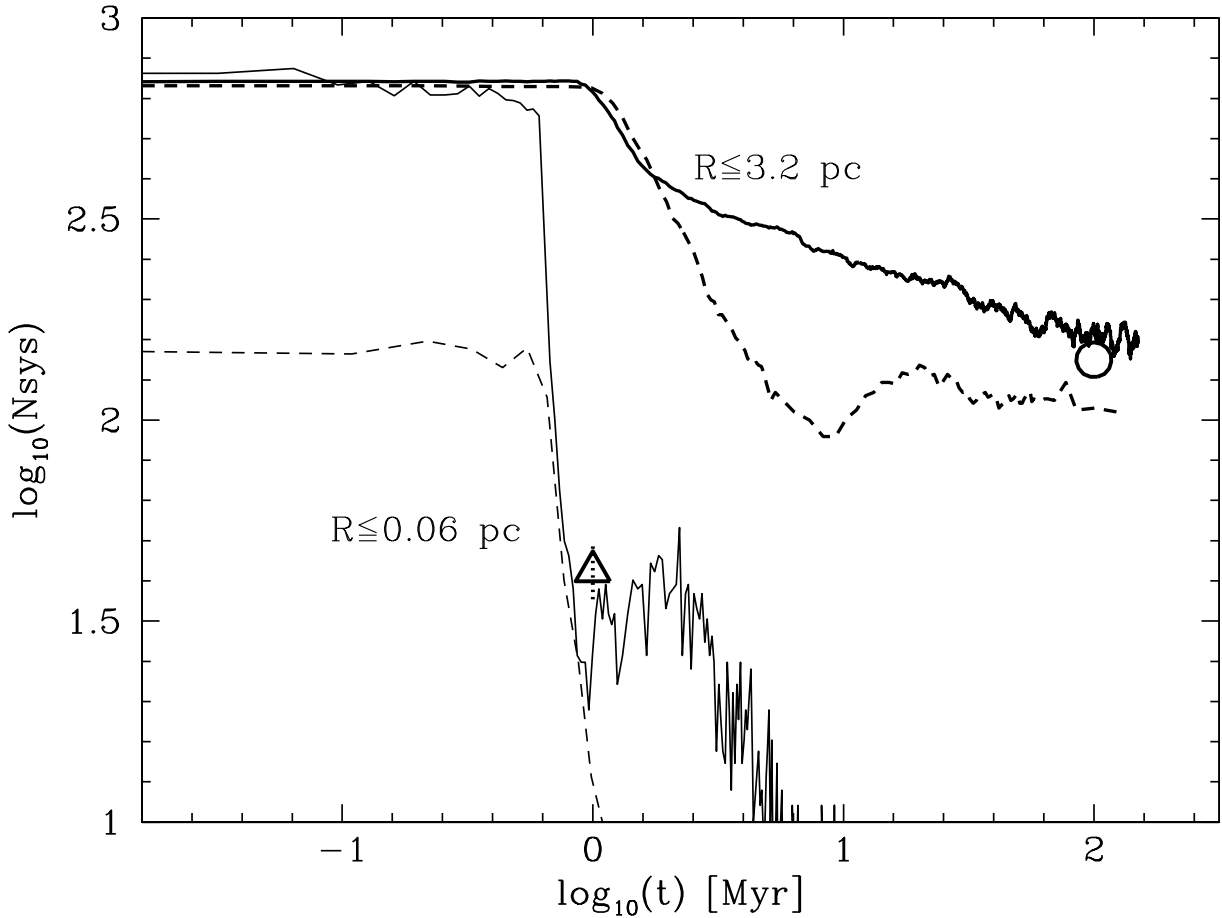}{15cm}{-90}{63}{63}{-250}{400}
\caption{The number of {\it systems} within $R\le3.2$~pc with primary
masses between 0.8~and 2.5~$M_\odot$ and binary companions with a
separation $a>130$~AU counted as separate stars (thick lines). The
corresponding datum for the Pleiades Cluster is shown as the open
circle (from Table~3 in Raboud \& Mermilliod 1998).  The thin lines
are the total number of {\it systems} within $R\le0.06$~pc. No binary
systems are resolved.  The triangle is the corresponding datum for the
Trapezium Cluster (McCaughrean \& Stauffer 1994). Dashed lines are for
model~A, and solid lines are for model~B.
\label{fig:ns}}
\end{figure}

\begin{figure}
\plotfiddle{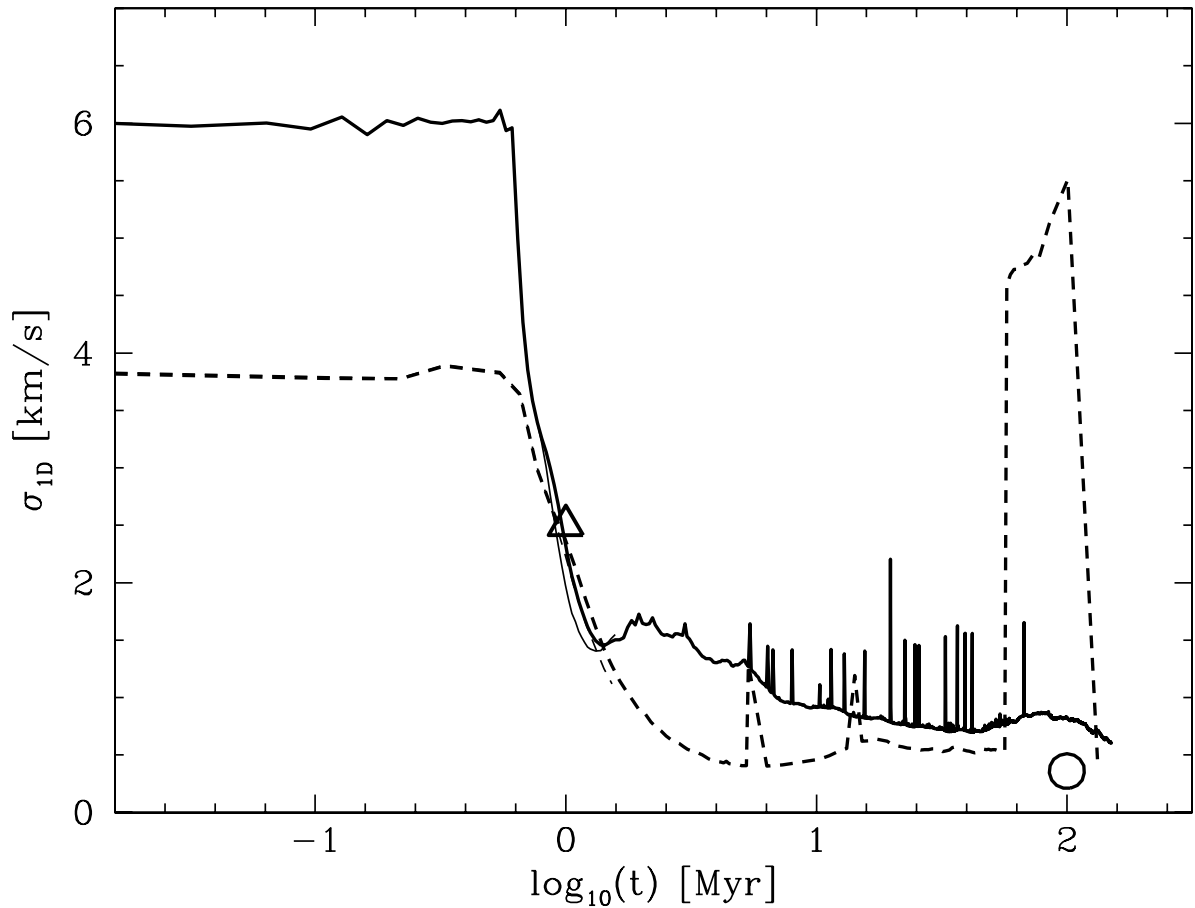}{15cm}{-90}{63}{63}{-250}{400}
\caption{The velocity dispersion of systems within $R\le3.2$~pc (thick
curves) and within $R\le2.5$~pc (thin curves). The value for the
Pleiades is the open circle (Raboud \& Mermilliod 1998) and the
triangle indicates the value for the ONC (Jones \& Walker 1988).  This
velocity dispersion is not corrected for the bulk radial expansion
(see Figs.~\ref{fig:vdo} and~\ref{fig:prv}).  Dashed lines are for
model~A, and solid lines are for model~B. The vertical excursions at
later times are due to energetic binary-star encounters, which eject
stars. Note in particular the large excursion between~60 and~100~Myr
in model~A. It is a result of the formation of a multiple system in
which the outer component has a high binding energy and thus a high
orbital velocity. The data reduction code only combines the innermost
binaries into one centre-of-mass system, and assumes the outer
companion to be a cluster field star.
\label{fig:vd}}
\end{figure}

The core radius is compared with observational constraints for the ONC
and the Pleiades in Fig.~\ref{fig:rc}.  Both models are in reasonable
agreement with the constraints.  Model~B, however, fits better near
$t=1$~Myr when Poisson fluctuations are still small. The same is true
for the number of systems shown in Fig.~\ref{fig:ns}. Comparison of
the available velocity dispersion measurements with the models also
leads to agreement (Fig.~\ref{fig:vd}).

\begin{figure}
\plotfiddle{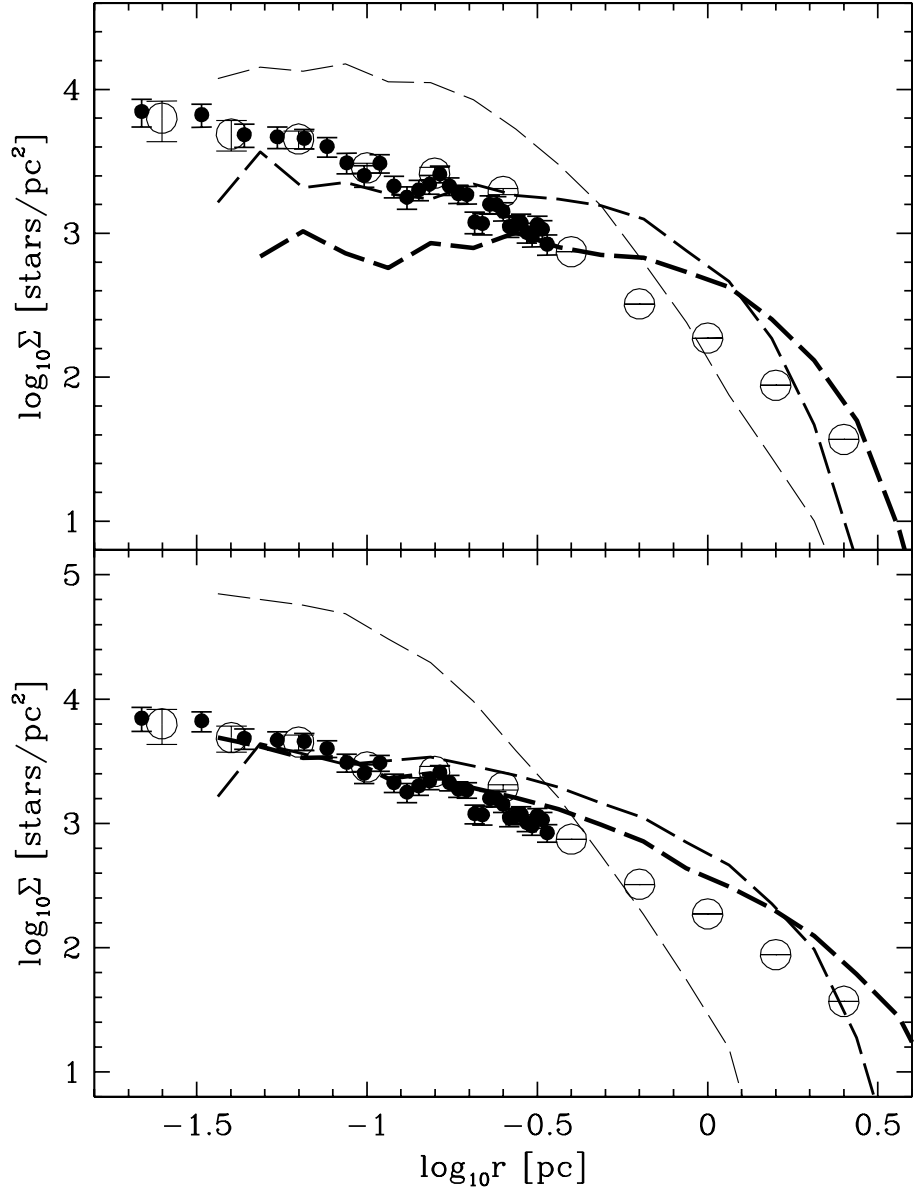}{15cm}{0}{64}{64}{-190}{-20}
\caption{The projected radial profile at $t=0, 0.87$ and 1.1~Myr (in
increasing thickness) for all {\it systems}. The open circles are
observational data from Hillenbrand (1997), and the solid circles are
from McCaughrean (1998, private communication). Upper panel is for
model~A, lower panel for model~B.
\label{fig:radpr_onc}}
\end{figure}

\begin{figure}
\plotfiddle{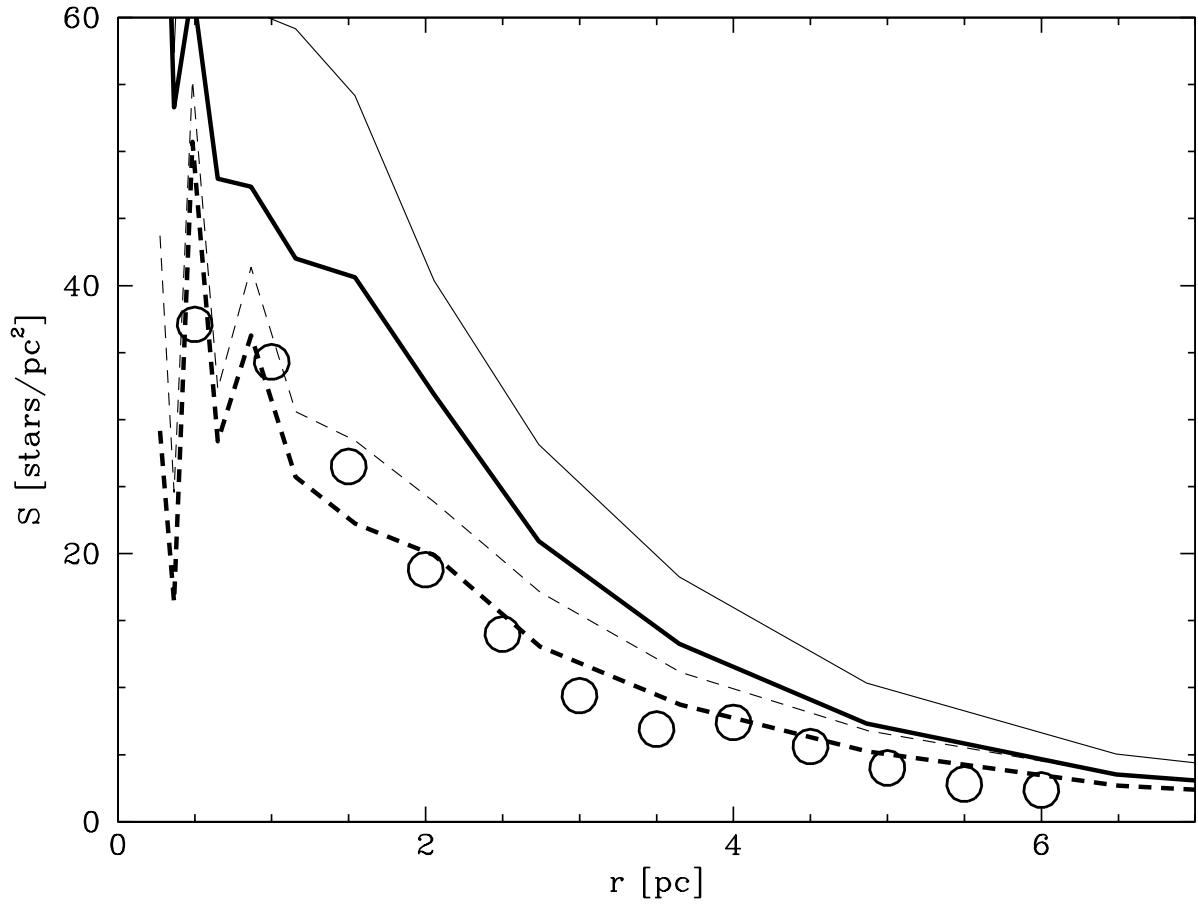}{15cm}{-90}{63}{63}{-250}{400}
\caption{The projected radial profile at $t=100$~Myr for {\it systems
in which at least the primary has $m\ge0.08\,M_\odot$} (thick curves),
and for all systems (thin curves).  The open circles are observational
data for the Pleiades from Pinfield, Jameson \& Hodgkin (1998, their
fig.~2). Dashed curves show model~A, solid curves model~B.
\label{fig:radpr_pl}}
\end{figure}

The projected radial density profiles are further constraints.
Ideally, the observed ONC and Pleiades profiles should be reproduced
by one model at $t\approx1$~Myr and $\approx100$~Myr, if the Pleiades
is a later rendition of the ONC. Model~A fails with the ONC
(Fig.~\ref{fig:radpr_onc}) in that the profile is too flat for
$r\simless 1$~pc, and too steep at larger radii, when
$t\approx0.9$~Myr. The model, however, fits the Pleiades rather well
(Fig.~\ref{fig:radpr_pl}).  Conversely, model~B fits the ONC
reasonably well for $t\approx1.1$~Myr (Fig.~\ref{fig:radpr_onc}), but
contains too many stars compared to the Pleiades at $t\approx100$~Myr
(Fig.~\ref{fig:radpr_pl}).

The deviations and agreements found here are very useful in discerning
what density profile and $N$ the ONC and Pleiades precursors are
likely to have had for a better reproduction of either cluster, and
perhaps both simultaneously. However, it would have to be considered
an impressive ``quirk'' of nature if it turns out that the ONC
actually {\it can} be viewed as a precursor of the Pleiades.
The present results have demonstrated that this is indeed the
case, to a surprising degree of accuracy. Further work will be needed
to study how much the various parameters (radial profile, $N$, IMF,
binaries, $\epsilon$) must be varied in order to get one model with
improved agreement for both clusters.

\subsection{Prediction for the bulk motion and velocity dispersion}
\label{sec:res:rv}
\noindent
If the ONC is the precursor of a Pleiades-like open cluster,
then it is expanding now.  The possible expansion of the Trapezium,
which is essentially the core of the ONC (Hillenbrand \&
Hartmann 1998), was already considered by KPM. In their model, which
assumed $\epsilon=0.42$ and a higher central number density but with
only 1600~stars, the projected bulk radial velocity $<\!\! v_{\rm
r}\!\!>=2.9$~km/s for stars with $r<0.41$~pc about 60000~yr after
instantaneous gas expulsion.

Models~A and~B can be evaluated to give the {\it predicted} expansion
rate of the ONC. The projected bulk radial motion, $<\!\! v_{\rm
r}\!\!>$, which is the average of the radial component of the
projected 2D velocity vectors of all stars with $r\le r_{\rm o}$ in
the observational plane, is plotted in Fig.~\ref{fig:prv}.  Model~A
gives an expansion velocity of $<\!\!  v_{\rm r}\!\!>=2.4$~km/s
($r_{\rm o}=2.5$~pc) and $<\!\!  v_{\rm r}\!\!>=0.6$~km/s ($r_{\rm
o}=0.41$~pc) at an age $t=0.9$~Myr (log$_{10}t=-0.05$), whereas
model~B expands with $<\!\! v_{\rm r}\!\!>=2.7$~km/s ($r_{\rm
o}=2.5$~pc) and $<\!\!  v_{\rm r}\!\!>=0.4$~km/s ($r_{\rm o}=0.41$~pc)
at the same age. The bulk radial velocity drops when the fastest stars
leave the field of view.

The line-of-sight velocity dispersion, $\sigma_{\rm los}$, decays at a
different rate after gas expulsion than the projected velocity
dispersion, because fast moving stars remain in the observational
field for a longer time along the line of sight.  Thus a difference
between the two is expected. This is shown in Fig.~\ref{fig:vdo},
where the line-of-sight and the projected 1D velocity dispersions are
compared. As expected, $\sigma_{\rm los}>\sigma_{\rm 1D,c}$, where
$\sigma_{\rm 1D,c}$ is the projected velocity dispersion corrected for
the bulk radial motion.  This is the velocity dispersion obtained from
ground-based proper motion studies, which apply special
photographic-plate-reduction techniques to eliminate distortions which
remove bulk rotation and bulk radial motions (see e.g. Jones \& Walker
1988; this is also discussed in KPM). For model~A at $t=0.9$~Myr,
$\sigma_{\rm los}=2.6$~km/s and $\sigma_{\rm 1D,c}=2$~km/s for
$r\le2.5$~pc and $r\le0.41$~pc. For model~B, $\sigma_{\rm
los}=2.8$~km/s and $\sigma_{\rm 1D,c}=2$~km/s for $r\le2.5$~pc with
smaller values if the measurement is confined to $r\le0.41$~pc.

The planned astrometric satellite missions DIVA (R\"oser 1999) and GAIA
(Lindegren \& Perryman 1996; Gilmore et al. 1998) should verify these
predictions of $<\!\!v_{\rm r}\!\!>$ for different $r_{\rm o}$.

A further generic prediction of these models is that the expanding
population of unbound stars (about $2/3$ of $N$ here) forms, at the
age of the Pleiades, a distinct co-eval group with an extension of
roughly 1~kpc ($\sigma_{\rm 3D}\times100\,{\rm Myr}\times2/\sqrt{3}$,
$\sigma_{\rm 3D}\approx9$~pc/Myr from Table~\ref{tab:mods}) and
co-moving with the Pleiades. This moving group, which we refer to as
'group~I', differs from the classical moving 'group~II' that forms
later from the long-term evaporation of Pleiades members through
two-body relaxation. The difference is kinematical in nature, in that
members of group~II diffuse away from their parent cluster mostly
through the two Lagrange points with a small relative velocity of
typically $\simless 1$~km/s (Terlevich 1987; de La Fuente Marcos 1997;
Portegies Zwart et al. 2000). The planned astrometric missions will be
able to distinguish these two moving groups associated with the
Pleiades, in which case this cluster-formation theory will be
ascertained. Group~I may in fact have already been observed as the
'Pleiades supercluster' by Eggen (e.g. Eggen 1998).

\subsection{The binary population}
\label{sec:res:binpop}
\noindent
The distribution of periods of binary systems measured in the Pleiades
(and the ONC) are not the birth distributions, since these have been
changed through dynamical interactions (KPM). The dynamical models
studied here allow this evolution to be quantified. The most obvious
hypothesis that one would consider testing is whether the Pleiades and
ONC binary populations are consistent with the same populations seen
in Taurus--Auriga, i.e. that the birth binary properties do not depend
on environment, apart from disruption through crowding and the cluster
tidal field.  Thus, models~A and~B assume Kroupa's (1995b, eq.~8)
birth distribution, which is essentially the Taurus--Auriga period
distribution of pre-main sequence binaries, and in this section the
evolved distributions are compared to the observational constraints
available for the ONC and the Pleiades. The stellar-dynamical
interaction between a primordial binary population and a young cluster
is described in some detail in Kroupa (2000b).

The evolution of the binary proportion of primaries in five mass
ranges for stars remaining in the cluster ($R\le3.2$~pc) is shown in
Fig.~\ref{fig:f}. Initially, the binary proportion of the massive
stars, $f_{\rm O}$, is reduced through disruption from crowding in the
dense clusters and through interactions, but it increases as the
massive stars gain new companions. A topic for future study will be
the comparison of the period and mass-ratio distributions of the
massive stars as a function of time with observational
constraints. For example, Mason et al. (1998) find a very high
proportion of binaries among massive stars, as is obtained in the
present calculations by $t>2.5$~Myr in model~A, and $t>16$~Myr in
model~B.  For the massive binaries, $f_{\rm O}$, Preibisch et
al. (1999) find that in the ONC the O~and~B primaries have on
average~1.5 companions. A more detailed analysis of the model data
will have to be performed to quantify the number of triple and
quadruple systems among the massive stars, before a final conclusion
concerning the primordial characteristics of massive binaries can be
made.

At the other extreme, BD binaries are more easily disrupted at an
early stage owing to their weaker binding energy, and by
$t\approx1$~Myr, $f_{\rm BD}\approx0.35$ (model~A) and 0.2~(model~B).
Overall, the Pleiades and ONC observational constraints for late-type
binaries are in acceptable agreement with the models.

The distribution of periods for late-type stellar
($0.08-1.5\,M_\odot$) and BD primaries is shown in Fig.~\ref{fig:p1}
at $t=0.9$~Myr and Fig.~\ref{fig:p100} at $t=100$~Myr. The
distributions in the former figure are compared with the initial
(after pre-main sequence eigenevolution) distributions, and with
observational constraints available for the ONC. By $t=0.9$~Myr
substantial depletion of the period distribution has occurred in both
models. However, only model~B leads to agreement with the measured
proportion of binaries with periods in the range
$P\approx10^5-10^{6.5}$~d, model~A retaining too many binaries with
these periods. Model~B is also consistent with the absence of
long-period ($P>10^7$~d) binaries in the ONC (Scally, Clarke \&
McCaughrean 1999).

The models are compared with the Pleiades constraints in
Fig.~\ref{fig:p100}, where a comparison is also made between the
period distributions for stars inside and outside the cluster. Both
models are consistent with the Pleiades data, although an intermediate
model in terms of central density would lead to improved
agreement. 

The distributed population of stars formed after gas expulsion (moving
group~I, Section~\ref{sec:res:rv}), has a higher binary proportion
than the stars remaining in the cluster.  This is evident in
Fig.~\ref{fig:p100} through the larger number of orbits at
$P>10^5$~d. In model~A there are slightly too many field binaries
relative to the Galactic field with periods in the range
$P=10^5-10^7$~d, although the disagreement is marginal.  Model~B
produces too few field binaries with $P\ge10^7$~d. It is thus somewhat
doubtful if either model can produce a Galactic field population, but
it is to be noted here that unlike the empirical G-dwarf
Galactic-field period distribution, the model period distributions do
not measure triple and quadruple systems. The data reduction performed
only finds the closest bound pairs, so that a third long-period
companion is missed.  This will be the subject of a future study.

In both figures, the BD binaries have a significantly depleted binary
proportion over all periods, which is a result of their small binding
energy.

A comparison of Figs.~\ref{fig:p1} and~\ref{fig:p100} shows that the
birth distribution of orbital periods in the ONC could have been
similar to that in Taurus--Auriga, which also leads to the Galactic
field, if the ONC began with a central density between that assumed in
models~A and~B.  The same conclusion holds true for the
Pleiades. Thus, the observational constraints do not exclude the
hypothesis that the star formation conditions in the ONC and the
Pleiades led to the same distribution of binary periods as in
Taurus--Auriga. That is, the initial period distribution function may
be universal.

\subsection{The mass function}
\label{sec:res:mf}
\noindent
There is much interest in the IMF as it is one of the most important
constraints on star-formation theories, as well as being the
fundamental quantity entering many astrophysical problems.  It is thus
useful to study the changes in the stellar and system mass function
(MF) as the clusters evolve, to obtain an insight into the sort of
changes that make the quasi-observed MF (i.e. the system MF) differ
from the IMF.

Pre-main sequence eigenevolution leads to small deviations of the IMF
from eq.~\ref{eq:imf} shown by the solid dots in Fig.~\ref{fig:mfn09},
because of the mass-gain of the secondary in close binary
systems. This is evident in Fig.~\ref{fig:mfn09}: the stellar IMF
(thin dashed histograms) contains fewer BDs and more massive
stars. The deviations are small and consequently we refer to both, the
thin dashed histogram and eq.~\ref{eq:imf}, as the IMF.

Assuming random pairing and $f_{\rm tot}=1$ gives the system MFs,
which are plotted at $t=0$ (thin solid histograms) and $t=0.9$~Myr
(thick solid histograms). At $t=0$, the system MF has a broad maximum
near $m=0.35\,M_\odot$. It falls below the stellar IMF by a factor of
about~4 at the hydrogen burning mass limit and has $\alpha_1=+0.6$
(instead of~$+1.3$) between~$0.08$ and~$0.5\,M_\odot$. It falls below
the IMF by an order of magnitude in the BD regime with $\alpha_0=
-0.8$ (instead of $+0.3$).  This difference is reduced by $t=0.9$~Myr
because most of the soft binaries have been disrupted, but significant
discrepancies persist (this is studied in greater depth in Kroupa
2000c).

Mass segregation becomes evident through a surplus of massive stars in
the central region. That this is not the case in model~A by
$t=0.9$~Myr is shown in the upper panel of
Fig.~\ref{fig:mfn09c}. However, by virtue of the significantly shorter
initial $t_{\rm cross}$, model~B has evolved much more before gas
expulsion, and shows significant mass segregation (lower panel in same
figure). The binary proportion has also decreased much more
(Fig.~\ref{fig:f}), so that the system and single star MFs are quite
similar.

Similarly, Fig.~\ref{fig:mfn100} shows the IMF for stars and systems
in comparison to the stellar and system MFs at $t=100$~Myr for stars
with $R\le15$~pc, which is approximately $R_{\rm tid}$ of the model
open cluster. Stellar evolution has removed all stars with
$m\simgreat10\,M_\odot$. The system MF falls significantly below the
stellar MF for $m<0.25\,M_\odot$ in model~A, by virtue of the larger
surviving binary proportion. In model~B (lower panel) the smaller
binary proportion implies that the system MF (thick solid histogram)
lies below the stellar MF (thick dashed histogram) only by an
approximately constant factor of~2 for $m<0.4\,M_\odot$.
Table~\ref{tab:100Myr_15pc} summarises the census of systems with
$R\le15$~pc.

The central ($R\le2$~pc) cluster region shows significant mass
segregation at $t=100$~Myr in both models (Fig.~\ref{fig:mfn100c}),
although it is much more pronounced in model~B (lower panel). Matching
the stellar MFs (dashed histograms) to the IMF (solid dots) near
$1\,M_\odot$, it can be seen that the central region has been
relatively depleted in BDs.  Table~\ref{tab:100Myr_2pc} summarises the
census of systems with $R\le2$~pc.

Comparison to observational estimates of the MF for the Pleiades is
not the main topic of this paper, but will be a focus in a future
study. Kroupa (1995c) showed that the Pleiades LF is consistent with
the Galactic field IMF, but those cluster models did not incorporate
gas expulsion nor massive stars.  At present, it suffices to state
that the {\it system} MF presented in Fig.~\ref{fig:mfn100} has
$\alpha_0=0.0$ for both models, i.e.  the slope is similar to the
actual IMF ($+0.3$). This is quite consistent with the estimate by
Martin et al. (2000) ($\alpha_0\approx0.53$), and the number of BD
candidate systems found (34), may be consistent with the present
models (Tables~\ref{tab:100Myr_15pc} and~\ref{tab:100Myr_2pc}) since
their survey area extends to $r\approx8$~pc for a Pleiades distance of
120~pc and they are not likely to have found the least massive BDs.
The overall shape of the {\it system} MF is also similar to the
observed MF (Meusinger, Schilbach \& Souchay 1996, their fig.10;
Hambly et al. 1999, their fig.11).

\section{DISCUSSION}
\label{sec:disc}
\noindent 
In the previous section it was seen that models~A and~B both reproduce
the ONC {\it and} Pleiades to a remarkable degree, suggesting that the
Pleiades may have looked similar to the ONC about~99~Myr
ago. Additional calculations with {\sc GasEx} but without a background
gas potential, i.e. assuming $\epsilon=1$, but with stellar evolution,
demonstrate that, for the Pleiades to have its present mass and
structure, $R_{\rm C}\approx1$~pc and $N\approx3000$ at birth.  Such
models are studied in detail by Portegies Zwart et al. (2000).  An
initial model in virial equilibrium without gas and as concentrated as
the ONC {\it can never} evolve to have the low concentration (central
number density $\rho_{\rm C}\approx11$~stars/pc$^3$, Pinfield, Jameson
\& Hodgkin 1998) which the Pleiades has now.  It is natural to assume
that the large core radius is a result of gas expulsion, and this
applies to all known nearby open clusters.

However, before the explanation presented here of how open clusters
form can finally be accepted as being correct, the bulk radial motion
predicted by these models (Fig.~\ref{fig:prv}) have to be verified
through observations. The ONC {\it must be expanding} with $<\!\!
v_{\rm r} \!\!>\approx2-3$~km/s for stars with $r\le r_{\rm
o}=2.5$~pc.  The expansion velocity should be smaller for a smaller
$r_{\rm o}$.  This will be an important goal of the upcoming
astrometric space observatories, DIVA (R\"oser 1999) and GAIA
(Lindegren \& Perryman 1996; Gilmore et al. 1998). If these satellites
can measure such a bulk expansion of the ONC, and detect the Pleiades
moving groups~I and~II (Section~\ref{sec:res:rv}), then the claim that
star-cluster formation has been solved can be made.

\subsection{Cluster formation despite instantaneous expulsion of more
than 50~\% gas mass}
\label{sec:disc:clform}
\noindent
Why does a {\it substantial} bound cluster form despite an sfe of
$\epsilon\approx0.3$?

The underlying reason is very simple and has been realised for some
years when the first expanding computations were done with
{\sc Nbody5}. In KPM the formation of a bound cluster was noted
although the expectation was an expanding OB association.  

In any stellar-dynamical system, the velocity distribution of the
stars has a tail with radial velocities near~0. In the event of the
expulsion of a large part of the mass, those stars which have turned
around on their orbits remain bound.  Moreover, the richness of the
emerging cluster depends on the number of stars in the velocity
distribution that have a velocity, $v$, smaller than the escape
velocity, $v_{\rm esc}$, from the system {\it after} the mass
(gas$+$unbound stars) is expelled. After mass expulsion, the
population with initially $v<v_{\rm esc}$ has a non-equilibrium
distribution of velocities, and as this distribution relaxes on the
new dynamical time-scale, more stars will be lost. The final
population of the cluster is therefore expected to depend sensitively
on the violent relaxation process, and thus on the accurate treatment
of close encounters as well as on the tidal field. For this reason
{\sc GasEx} was developed during the subsequent years~1998--1999 with
additional improvements in the underlying {\sc Nbody6} code.

If the requirement of detailed predictions is relaxed, important
insights into the above cluster-forming process can be gained through
analytical work.  Adams (2000) considers the stellar distribution
function, which is a solution of the collisionless Boltzmann equation
(CBE), in a potential that is a combination of stars and gas. Use of
the CBE, and thus neglect of near-neighbour interactions, should be a
reasonable approximation, as long as the crossing time of the cluster,
which sets the time-scale for stellar interactions, is longer than the
gas-expulsion time-scale.  The distribution function gives the
fraction of stars that remains bound after the gas is removed
instantaneously.  Adams finds that for $\epsilon=0.3$ (the case
assumed here) between ${\cal F}=35$ and~70~\% of the stars in the
initial cluster have velocities smaller than the escape velocity after
gas removal. This work shows that the central concentration and a
velocity anisotropy that is predominantly radial in the outer initial
cluster regions lead to more massive remnant clusters.

However, in Adams' models, the gas distribution is more extended than
that of the stars, so that the effective $\epsilon$ is larger within
the stellar cluster, being as high as~0.9 (Geyer \& Burkert 2000).
Furthermore, the estimates obtained using this approach do not take
into account the additional stars lost during the relaxation process
to new dynamical equilibrium, nor the Galactic tidal field, and so
constitute upper limits on ${\cal F}$ without knowledge of the lower
limits, which are the relevant ones for the problem. 

It may thus not be surprising that Adams' conclusions deviate
substantially from those of Geyer \& Burkert (2000), who employ a
collisionless $N$-body code and adopt the same density distribution
for the gas and stars, and show that {\it no cluster forms} if
$\epsilon\le0.40$ for instantaneous gas expulsion. An external
Galactic tidal field (not applied by Geyer \& Burkert) will further
increase the critical $\epsilon$ needed for cluster formation in the
collisionless regime.  The computations with {\sc GasEx}, on the other
hand, result in the formation of {\it substantial clusters} containing
about $1/3$ of the initial number of stars, despite $\epsilon=0.3$,
despite applying a {\it Galactic tidal field}, and despite evolving
stars, which together worsen the survival chance of an expanding
system.

The reason for the formation of substantial clusters in our
computations are {\it gravitational interactions between neighbours
during the radially expanding flow}, which are not correctly treated
in the collisionless approximation that underlies all hitherto
available work. This mechanism leads to the emergence and
amplification of a non-radial velocity dispersion even in the event of
an initially purely radial outwards flow, and it operates most
efficiently for the stars populating the slow tail of the velocity
distribution. Kinetic energy is thus redistributed from the radial
flow into orbital motions with non-zero angular momentum about the
origin of the expanding flow, allowing a substantial part of the
stellar system to condense as a bound entity, thereby defining and
filling it's tidal radius. Animated additional computations with {\sc
GasEx} sometimes show {\it multiple clusters} emerging from the radial
flow, which, if massive enough so that two-body relaxation does not
evaporate them too fast, survive and merge with the dominating
sub-cluster to form the Galactic cluster. Essentially, the same
mechanism operates in the inverse problem of a finite-$N$ system that
collapses from rest. The increasing non-radial velocity dispersion
that results from the grainy potential limits the maximum collapse
factor in dependence of $N$, as has been studied in detail by Aarseth,
Lin \& Papaloizou (1988). Clearly, much work remains to be done along
similar lines in the case of a radial outwards flow.

\subsection{Comparison with similar work}
\label{sec:disc:simw}
\noindent
That a bound ``core'' remains despite substantial mass-loss has also
been noted by Lada et al. (1984) in their pioneering computations with
$\le100$~stars. The models considered here correspond to the case
$\tau_{\rm R}<<t_{\rm cross}$ ($\tau_{\rm R}$ being the gas-removal
time) in their fig.2, although they varied the radius of the
background gas, $R_{\rm pl,g}$, in contrast to the variation of the
mass, $M_{\rm g}$, applied here.  The Galactic tidal field is not
incorporated in their investigation.

By scaling their results to a Pleiades-like cluster, they find that
it's precursor must have had a central density $\rho_{\rm
C}\approx10^{5.2}$~stars/pc$^3$ assuming a mean stellar mass
$<\!\!m\!\!>=0.4\,M_\odot$ (their table~2 for $\tau_{\rm R}=0$ and
$\epsilon=0.4$). This is close to the actual central density {\it
observed} in the Trapezium, and is intermediate to that assumed in
models~A and~B here, which, however, assume $\epsilon=0.3$ (note that
Geyer \& Burkert 2000, who revisit the Lada et al. problem with a
collisionless code, find no cluster formation if $\epsilon=0.3$).

The high central density of the Trapezium (McCaughrean \& Stauffer
1994) was not known at the time of Lada et al. (1984), and they
concluded that such high densities are not realistic, because this has
not been observed in molecular clouds. They reason therefore that
Pleiades-like clusters must form from less dense initial conditions
with $\tau_{\rm R}>>t_{\rm cross}$. This can only be achieved in the
{\it absence of O stars}, a suggestion also made by Elmegreen (1983)
and Elmegreen \& Efremov (1997).  In view of this reasoning, a
possible precursor to a Pleiades-like cluster appears to be the
$\rho$~Oph cluster which contains no O~stars. However, today we know
that the stellar population in the $\rho$~Oph cluster is too
sparse. It amounts to about~100~stars, and the total mass in gas and
stars is about $500\,M_\odot$ (Zinnecker et al. 1993; Luhman \& Rieke
1999). In contrast, the Pleiades contains a stellar mass of
between~500 and~about~5000$\,M_\odot$ (Pinfield et al. 1998; Raboud \&
Mermilliod 1998).

The line of argument followed here thus differs from the conclusions
arrived at by Lada et al. (1984) and Elmegreen \& Efremov (1997),
because today we know that the very high central densities needed to
form a Pleiades-like cluster from the ONC are available in the
presence of O~stars, and that O~stars form in profusion in rich
clusters. The calculations presented here demonstrate that even in the
presence of O~stars, which lead to explosive gas ejection, a
substantial bound cluster remains.

The reaction of young clusters to mass loss is also studied by Goodwin
(1997a; 1997b) and Geyer \& Burkert (2000), but in the context of the
formation of globular clusters. Goodwin uses the {\sc Nbody2} code
with $N=1000$ particles, and applies a time-varying background
potential in much the same way as done here, but also studies the
effects of gas expulsion via an expanding shell.  The gas is removed
over a time-scale of a few Myr initially, and finally, after 10~Myr as
a result of supernovae. Bound clusters form for
$\epsilon\ge0.25$. Using a softened (``collisionless'') potential is,
in principle (but see Section~\ref{sec:disc:clform}), a reasonable
approximation for this purpose, because the evolution is restricted to
being shorter than a relaxation time, but the initially slow gas
depletion favours the formation of bound clusters, since the stellar
orbits can adjust adiabatically to the varying potential, until the
residual gas is blown out through supernovae. The results give useful
insights into the likely initial conditions of massive
clusters. 

However, observational evidence appears to imply that gas expulsion
occurs on a shorter time-scale. For example, the massive central R136
cluster in the 30~Doradus Nebula in the Large Magellanic Cloud has an
age younger than about 2~Myr (Selman et al. 1999), but is already
devoid of gas. In this respect, the collisionless computation by Geyer
\& Burkert (2000) are interesting, since they consider instantaneous
expulsion and find that no cluster forms unless
$\epsilon\ge0.4$. Their more realistic SPH treatment of the gas
arrives at the same results as with an analytical background
potential, lending credence to our approach.

\subsection{An ONC peculiarity}
\label{sec:disc:pec}
\noindent
One observational finding further supports the present result that the
Pleiades may have formed from an ONC-like object, namely the short
life-time of the circum-stellar features found around stars in the
Trapezium. Henney \& O'Dell (1999) measure the mass-loss rates of four
of these objects, and find that they could not have been exposed to
the UV flux from the central O~stars for longer than about
$10^4$~yr. These particular objects could be crossing the inner
cluster region after spending more time at larger radii. However, the
large proportion of stars with circum-stellar material does indicate
that the destructive irradiation may indeed have turned on relatively
recently. If this was the case, then it would appear natural to
associate this event with the gas-expulsion time.

\section{CONCLUDING REMARKS}
\label{sec:concl}
\noindent 
This contribution presents the first results that were obtained using
{\sc GasEx}, thereby also being the first time that a high-precision
$N$-body code that treats close stellar encounters accurately is
applied to the problem of star-cluster formation.  The results differ
to those obtained with 'collisionless' $N$-body codes and analytical
estimates, in that substantial clusters form despite a low sfe
($\epsilon=0.3$) and rapid gas-expulsion ($\tau_{\rm M}<<t_{\rm
cross}$) due to the embedded OB stars, and despite applying a Galactic
tidal field and mass-loss from evolving stars, which further limit the
survival chances for an emerging star cluster.

Following Kroupa's (2000a) identification of possible initial states
of the ONC, two calculations were performed of a binary-rich embedded
cluster with brown dwarfs. The extreme situation of very rapid gas
expulsion together with a large number of massive stars is studied in
order to investigate the likely fate of the ONC. 

The rather startling result is that a Pleiades-like cluster forms
containing about $1/3$ of the initial number of stars. This result is
startling because all previous and contemporary work leads to complete
cluster dissolution for such a low sfe, the reason being the neglect
of the collisional nature of a finite-$N$ system.

A possible (non-causal) connection between the ONC and the Pleiades is
thus established. This scenario suggests that Galactic clusters form
with relatively large core radii and by filling their tidal radii, as
a result of the expansion after gas loss from a compact state similar
to the ONC. The scenario also suggests that they form as nuclei of
expanding OB associations. The calculations show that the primordial
binary population in the ONC and the Pleiades could have been
indistinguishable from the Taurus--Auriga population, indicating that
the {\it initial period distribution function may be universal}. An
analytical description of this function is given by eq.~8 in Kroupa
(1995b).

\begin{figure}
\plotfiddle{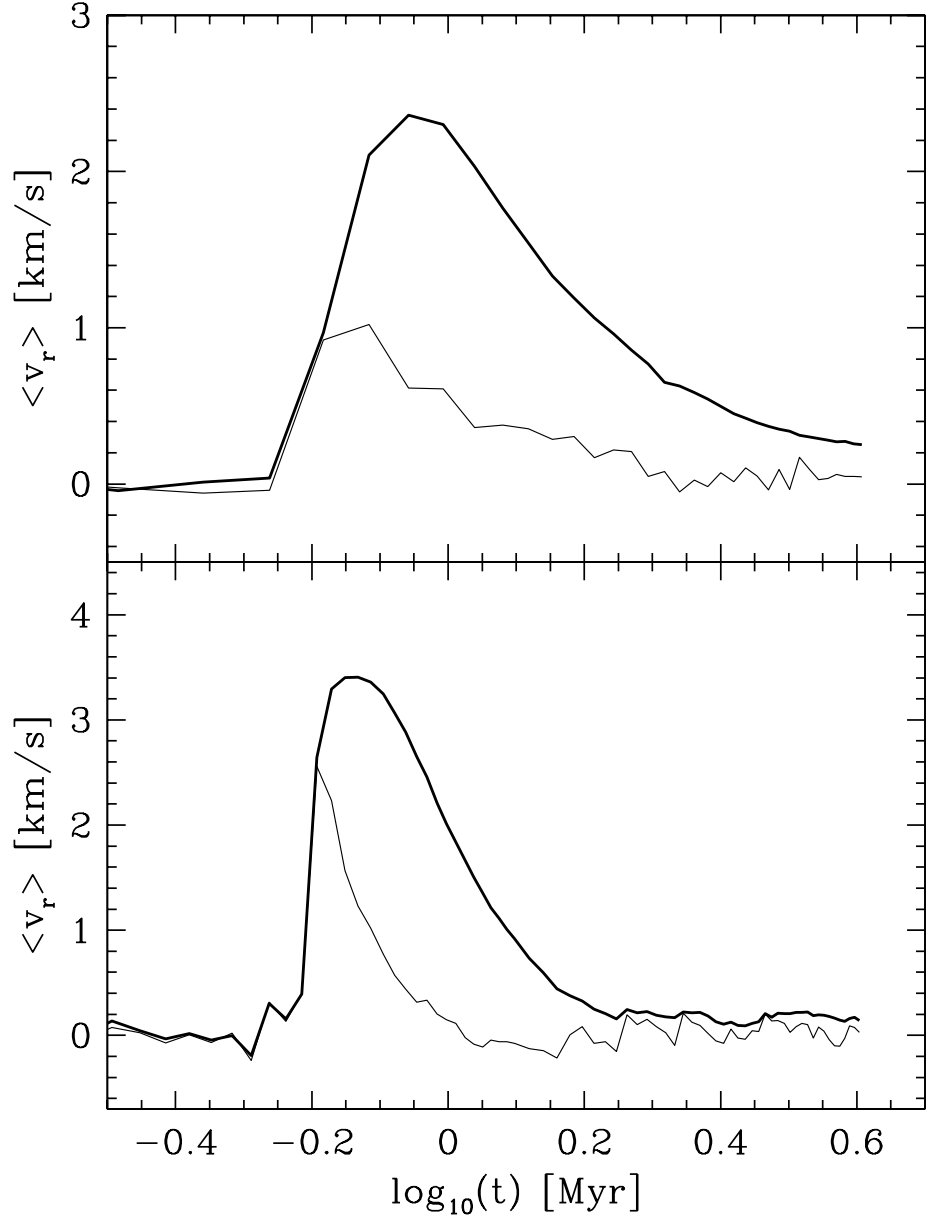}{15cm}{0}{64}{64}{-190}{-20}
\caption{Prediction of the projected bulk radial velocity of the ONC,
$<\!\!v_{\rm r}\!\!>$, measured for stars with $r\le2.5$~pc (thick
lines) and $r\le0.41$~pc (thin lines; cf. KPM). 
Upper panel is for model~A, lower panel for model~B.
\label{fig:prv}}
\end{figure}

\begin{figure}
\plotfiddle{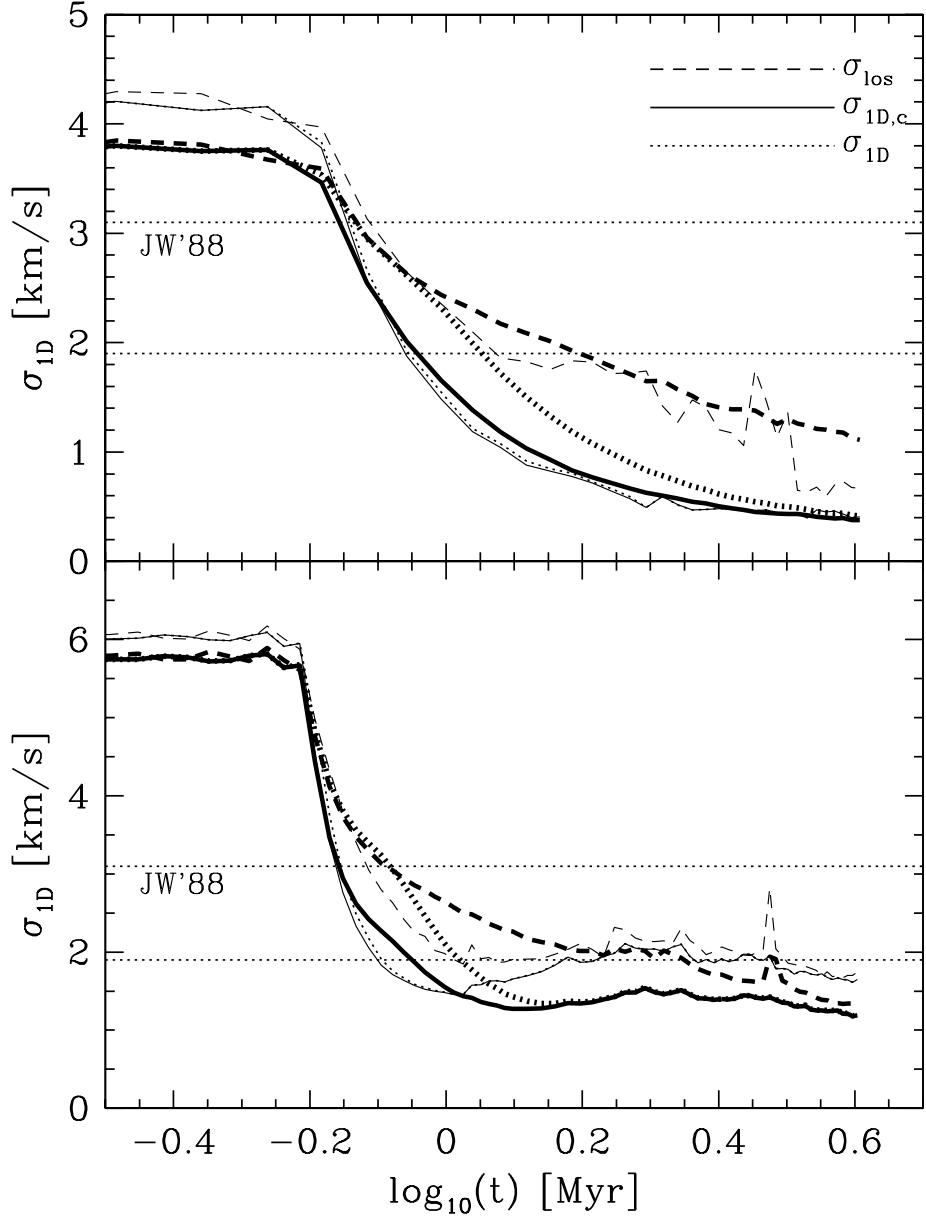}{15cm}{0}{64}{64}{-190}{-20}
\caption{ Line-of-sight (dashed curves) and the 1D velocity
dispersions in the observational plane corrected for the radial
expansion (solid curves) and uncorrected (dotted curves), for stars
with $r\le2.5$~pc (thick curves) and $r\le0.41$~pc (thin curves). The
three-sigma range for the projected (proper motion) velocity
dispersion of ONC stars is shown as the region between the horizontal
dotted lines (Jones \& Walker 1988). These authors find no significant
evidence for a radial variation.  The upper panel is for model~A and
the lower panel for model~B.
\label{fig:vdo}}
\end{figure}

\begin{figure}
\plotfiddle{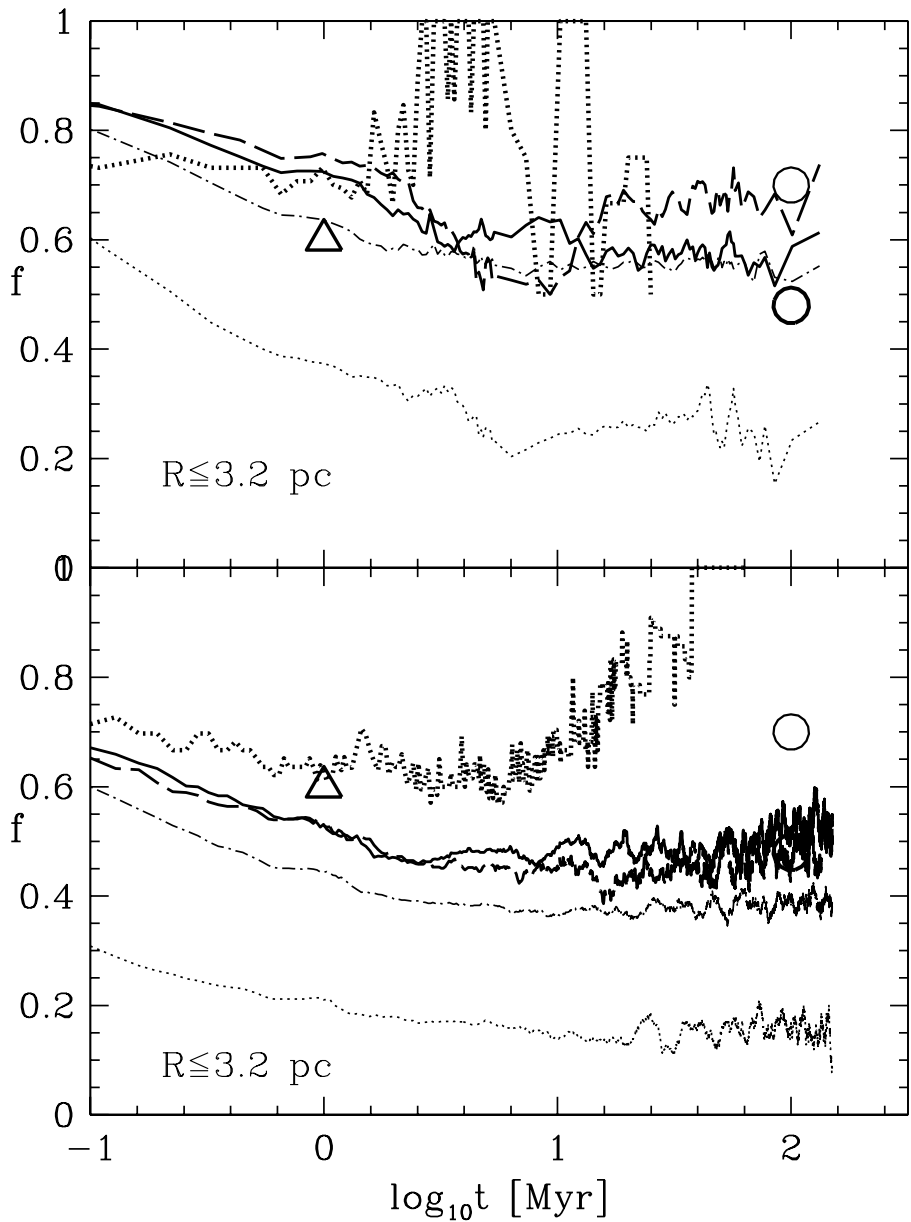}{15cm}{0}{64}{64}{-190}{-20}
\caption{The binary proportion (upper panel: model~A, lower panel:
model~B). The proportion of binaries with primary masses
$>8\,M_\odot$, $f_{\rm O}$ (thick dotted curve), $1-8\,M_\odot$,
$f_{\rm IM}$ (thick dashed curve), and $0.5-1\,M_\odot$, $f_{\rm K}$
(thick solid curve). M~dwarf primaries ($0.08-0.5\,M_\odot$) have a
binary proportion, $f_{\rm M}$ (dash-dotted line), whereas brown
dwarfs ($0.01-0.08\,M_\odot$), $f_{\rm BD}$, are shown as the thin
dotted line.  The open circles are the Pleiades binary proportion from
Raboud \& Mermilliod (1998) for primaries in the mass range
$0.8-2.5\,M_\odot$, approximately, whereas K\"ahler (1999) finds a
value as high as $f_{\rm tot}=0.70$ to be possible. The Raboud \&
Mermilliod value is a lower limit because not all faint companions are
detected.  Prosser et al. (1994) find a binary proportion for the
Trapezium Cluster similar to the Galactic field, which is indicated by
the open triangle.
\label{fig:f}}
\end{figure}

\begin{figure}
\plotfiddle{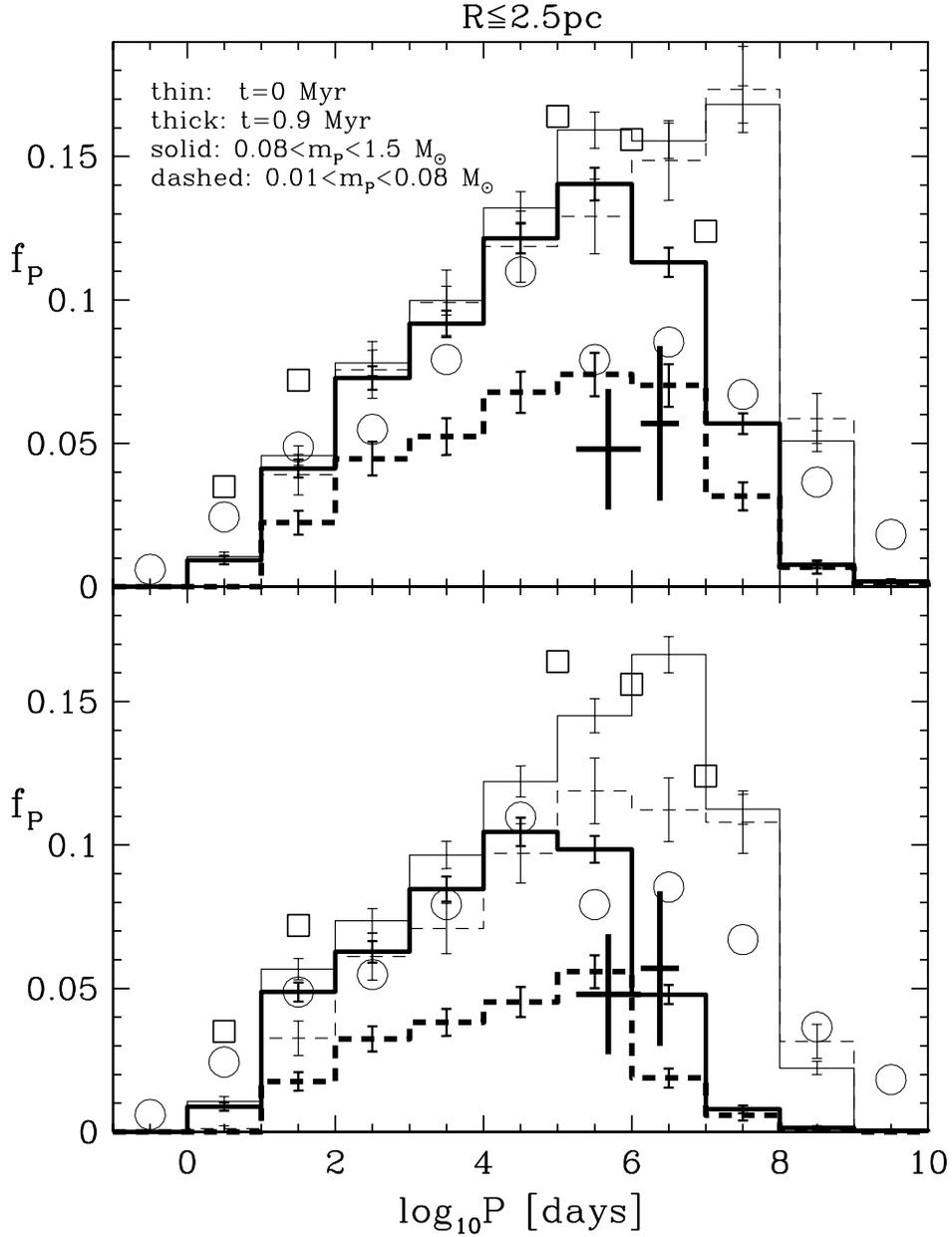}{15cm}{0}{64}{64}{-190}{-20}
\caption{The period distribution for late-type binaries within the
central 2.5~pc radius (upper panel: model~A, lower panel:
model~B). Thin histograms are the initial distributions for stellar
($0.08-1.5\,M_\odot$, solid histograms) and BD ($0.01-0.08\,M_\odot$,
dashed histograms) primaries. The distributions at 0.9~Myr are given
by the thick histograms.  Main-sequence G-dwarf multiple systems are
the open circles (Duquennoy \& Mayor 1991) and pre-main sequence
systems mostly in Taurus--Auriga are open squares (log$_{10}P>4$:
K\"ohler \& Leinert 1998, log$_{10}P=3.5$: Richichi et al. 1994,
log$_{10}P<2$: Mathieu 1994). The two large crosses at
log$_{10}P=5.69$ and log$_{10}P=6.38$ are ONC one-sigma observational
constraints for $r<0.3$~pc and $0.07<r<0.3$~pc, respectively (Petr
1998).
\label{fig:p1}}
\end{figure}

\begin{figure}
\plotfiddle{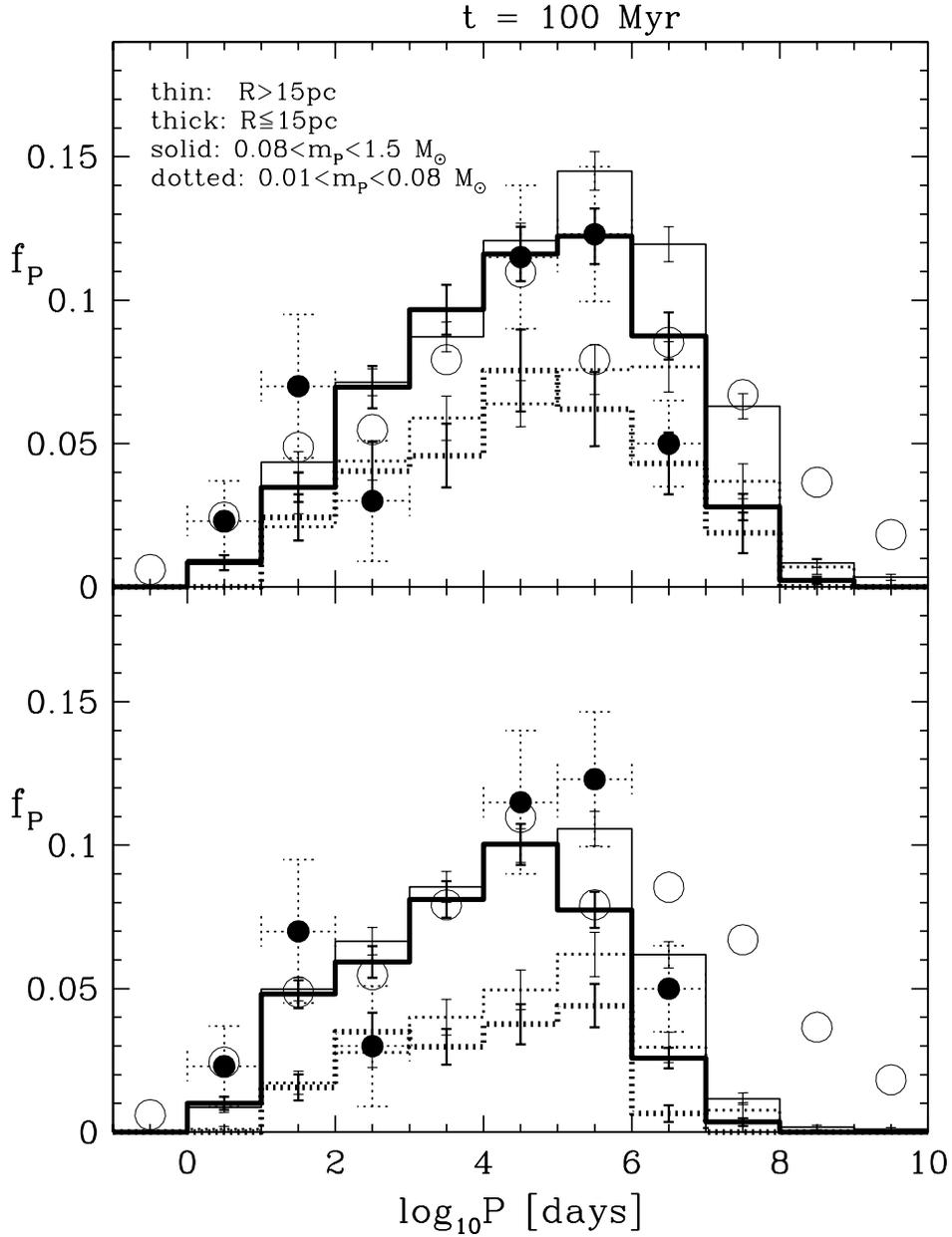}{15cm}{0}{64}{64}{-190}{-20}
\caption{The period distribution for late-type binaries within and
outside the central 15~pc radius at $t=100$~Myr (upper panel: model~A,
lower panel: model~B). Open circles are for Galactic field G-dwarf
systems (Duquennoy \& Mayor 1991). Galactic-field~K and~M dwarf
binaries have the same distribution (fig.~1 in Kroupa 1995a).  The
filled circles are observational constraints for Pleiades binaries
from Bouvier, Rigaut \& Nadeau (1997) (log$_{10}P>4$) and Mermilliod
et al. (1992) (log$_{10}P<3$).
\label{fig:p100}}
\end{figure}

\begin{figure}
\plotfiddle{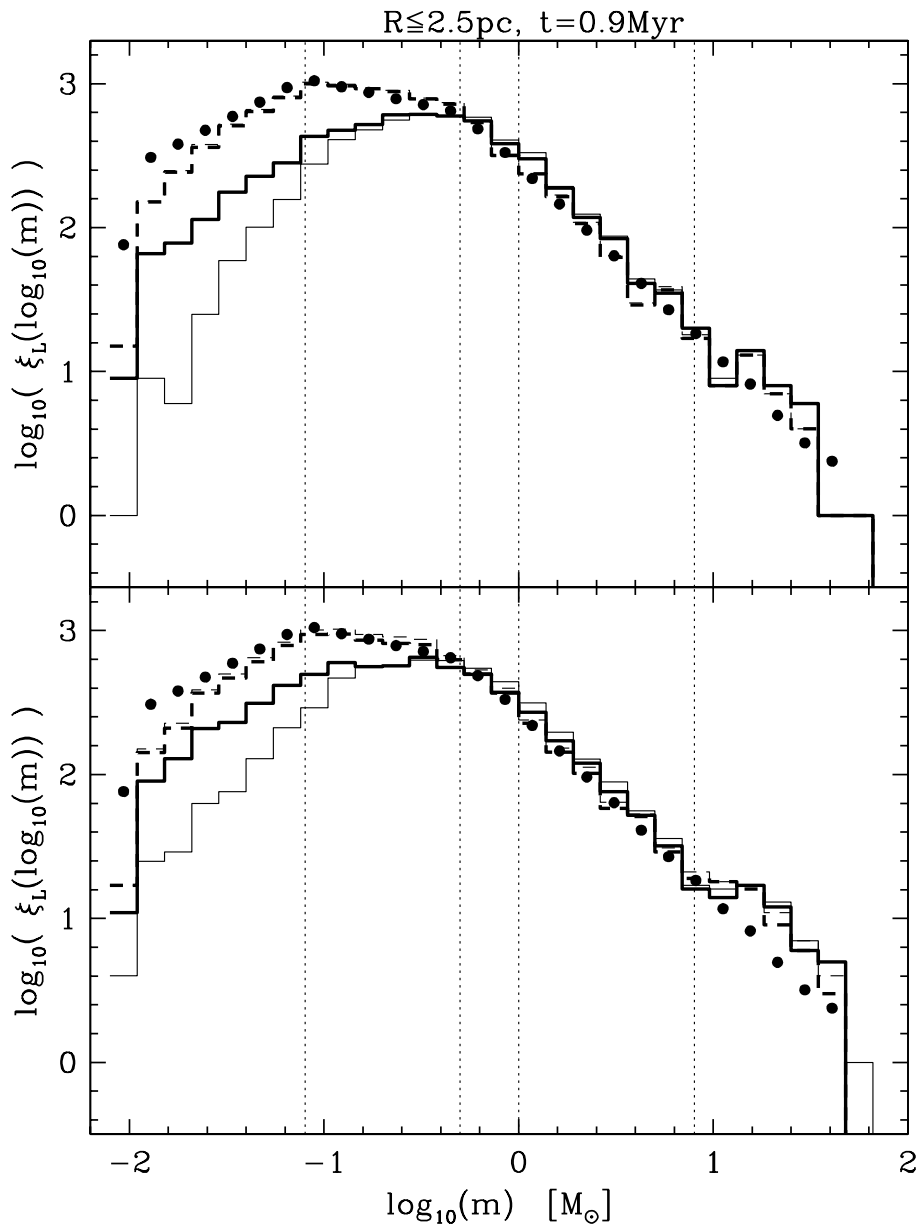}{15cm}{0}{64}{64}{-190}{-20}
\caption{The system (solid histograms) and stellar (dashed histograms)
mass functions initially ($t=0$, thin histograms) and at $t=0.9$~Myr
(thick histograms) within $R\le2.5$~pc, which is approximately the
radius of the Hillenbrand (1997) survey of the ONC. The solid dots are
eq.~\ref{eq:imf} with identical scaling in the upper and lower panels.
Slight deviations of the thin dashed histogram from eq.~\ref{eq:imf}
are due to the adopted {\it pre-main sequence eigenevolution} during
which binary secondaries gain mass in dependence of the initial
orbital semi-major axis (Section~\ref{sec:initpop}). The vertical
dotted lines delineate the masses at which the simplified stellar
types of Table~\ref{tab:100Myr_15pc} change, the two left-most also
being the masses at which $\alpha_i$ changes (eq.~\ref{eq:imf}).
Upper panel is for model~A and lower panel is for model~B.
\label{fig:mfn09}}
\end{figure}

\begin{figure}
\plotfiddle{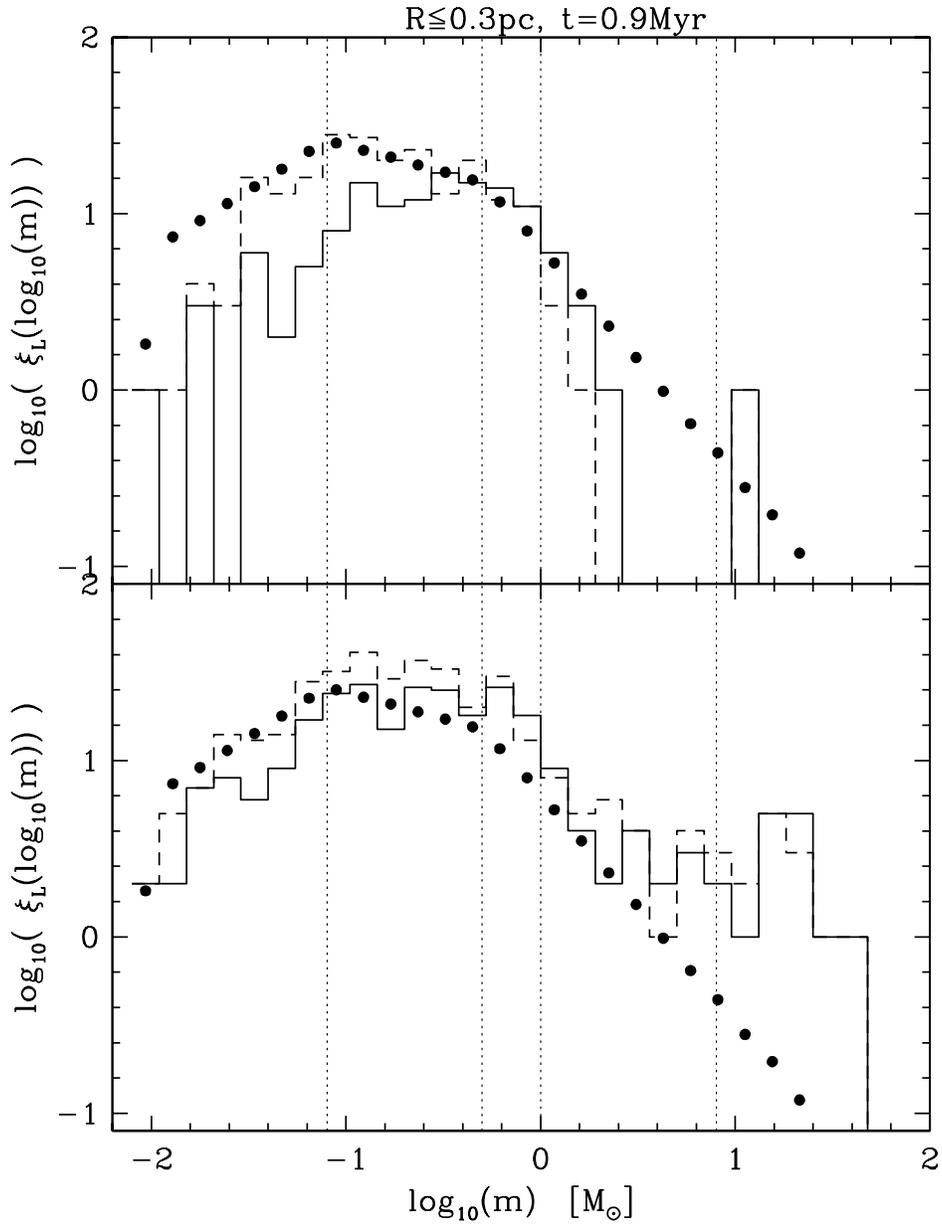}{15cm}{0}{64}{64}{-190}{-20}
\caption{The system (solid histograms) and stellar (dashed histograms)
mass functions at $t=0.9$~Myr within $R\le0.3$~pc, which encompasses
approximately the Trapezium Cluster. Otherwise as
Fig.~\ref{fig:mfn09}. The IMF (solid dots, eq.~\ref{eq:imf}) has the
same normalisation in both panels. Note the advanced mass segregation
in the lower panel. 
\label{fig:mfn09c}}
\end{figure}

\begin{figure}
\plotfiddle{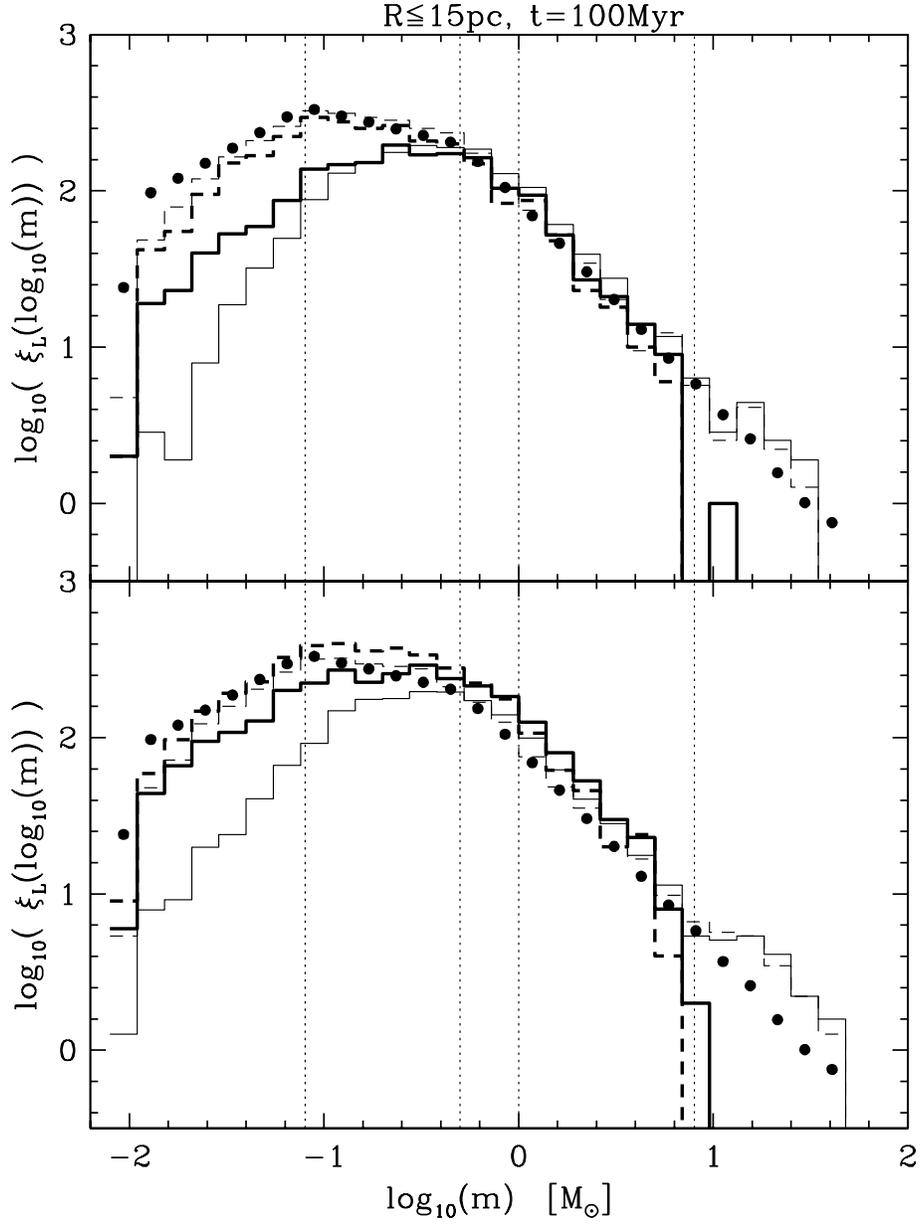}{15cm}{0}{64}{64}{-190}{-20}
\caption{The system (solid histograms) and stellar (dashed histograms)
mass functions initially ($t=0$, thin histograms, scaled to the thick
histograms: $\xi_{\rm L}(t=0)/3.16$) and at $t=100$~Myr (thick
histograms) within $R\le15$~pc, which is approximately the tidal
radius at $100$~Myr (Fig.~\ref{fig:lagr}). Otherwise as
Fig.~\ref{fig:mfn09}. The IMF (solid dots, eq.~\ref{eq:imf}) has the
same normalisation in both panels.
\label{fig:mfn100}}
\end{figure}

\begin{figure}
\plotfiddle{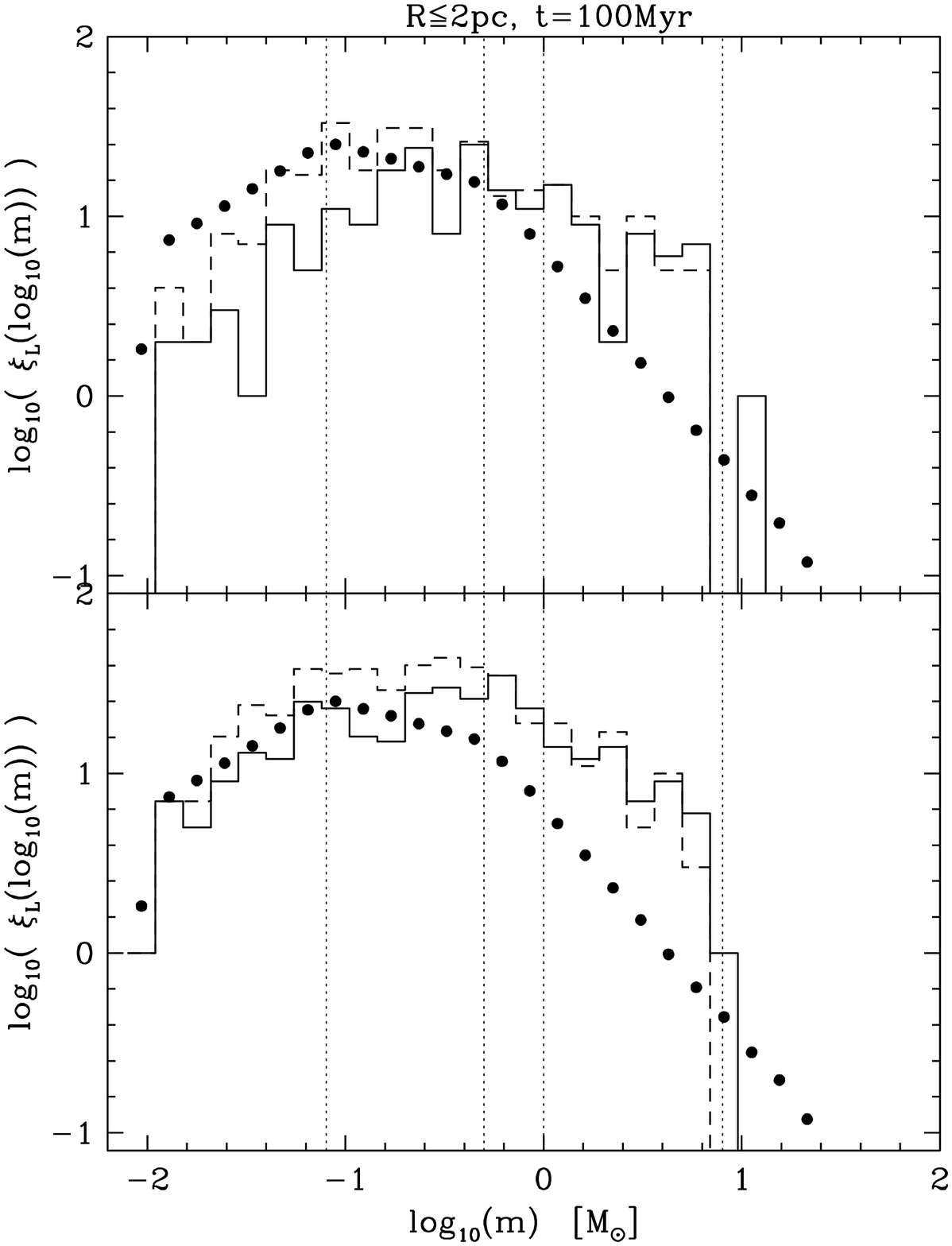}{15cm}{0}{64}{64}{-190}{-20}
\caption{The system (solid histograms) and stellar (dashed histograms)
mass functions at $t=100$~Myr within $R\le2$~pc, which encompasses the
inner region of the Pleiades Cluster (Fig.~\ref{fig:radpr_pl}).
Otherwise as Fig.~\ref{fig:mfn100}. The IMF (solid dots,
eq.~\ref{eq:imf}) has the same normalisation in both panels, which is
also identical to that in Fig.~\ref{fig:mfn09c}.
\label{fig:mfn100c}}
\end{figure}

The most important prediction of this scenario of cluster formation is
that the ONC must now be expanding, for it to be a precursor of a
Pleiades-like cluster.  The bulk expansion velocity and the
line-of-sight and projected (proper motion) velocity dispersions are
predicted in Figs.~\ref{fig:prv} and~\ref{fig:vdo}. This expanding
population forms a moving group (referred to here as 'group~I'), which
differs from the classical moving 'group~II' that results from secular
loss of cluster stars due to two-body relaxation.  Future astrometric
satellites will be able to confirm these predictions and differentiate
between groups~I and~II. Either way, a very important lesson about
cluster formation will be learned. An additional diagnostic of the
dynamical state of a young cluster is the radial profile of the binary
proportion, as stressed by KPM.

In this model of cluster formation, roughly $2/3$ of the initial stars
become field stars immediately after gas expulsion. These stars retain
a somewhat higher binary proportion than is seen in the Pleiades, and
form an expanding OB association and later moving group~I.  The
presence of wide binaries in the Galactic field poses important
constraints on the possible overall contribution to Galactic field
stars from such gas-expelled cluster stars (c.f. Scally et
al. 1999). The same applies to massive field stars and their
multiple properties. Clearly this poses a very rich field for further
study, and will improve our knowledge of the Galactic field
population.

The initial conditions assumed here (Table~\ref{tab:mods}) are only
two from the parameter space of possible solutions for an expanding
ONC constrained by Kroupa (2000a). There is some redundancy in the
parameters. Notably, a smaller $N$ can be compensated by a larger
$\epsilon$ and/or larger $\alpha_3$ (fewer massive stars) for the
Pleiades, and it will be important to delineate the allowed range of
these parameters for the Pleiades, Praesepe and Hyades clusters.
Collapsing models will also be studied to obtain predictions.

The findings presented here have important implications for the
formation of massive star clusters, some of which evolve to globular
clusters. Observations of young massive clusters, for example R136 in
the Large Magellanic Cloud, indicate that the gas is expelled
extremely rapidly and possibly before the first massive stars
explode. The observed haloes of stars (Elson, Fall \& Freeman 1987)
around some of the clusters suggest that a large proportion of the
stars formed in massive embedded clusters become unbound, much in the
way investigated here, and as also stressed by Goodwin (1997b).  A
significant proportion of Galactic spheroid (or 'stellar halo') stars,
that outnumber the stars in all globular clusters by at least 10:1
(Binney \& Merrifield 1998), may thus be stars that formed in
globular-cluster precursors but became unbound as a consequence of
gas-expulsion. The distribution of the dynamical properties (binarity,
and mass-ratio and period distributions) of the stars reflect such
events, and the stars that got unbound during gas expulsion should
have different phase-space properties than the stars lost from
globular clusters as a result of their long-term evolution in the
Galactic potential, for example by virtue of the general shrinkage of
the globular cluster system as a result of dynamical friction.

Finally, a more realistic variation of the present models would
include a primordial segregation of masses such that the most massive
stars are located in the core of the embedded cluster. In the event of
gas expulsion, the less massive stars located at larger radii would
leave the cluster preferentially. This scenario implies that the MF in
young clusters such as the Pleiades might be depleted in low-mass
stars and BDs relative to their IMF, so that the study of the IMF via
such young clusters would be compromised. Future work with {\sc GasEx}
will address this issue.

\acknowledgements 
\vskip 10mm
\noindent{\bf Acknowledgements}
\vskip 3mm
\noindent
PK acknowledges support through DFG grant KR1635.

%

\clearpage


\begin{table}
{\small
\begin{minipage}[t]{9cm}
\vskip 5mm

\begin{center}
\begin{tabular}{ccccccccccccc}

model &$N_{\rm sing}$ &$N_{\rm bin}$ &$R_{0.5}$ &$<\!\!m\!\!>$ 
&$\sigma_{\rm 3D}$ &$t_{\rm cross}$ &log$_{10}\rho_{\rm C}$
&$M_{\rm st}$ &$M_{\rm g}$ &$R_{\rm 0.5,{\rm g}}$ 
&$\tau_{\rm M}$ &$t_{\rm D}$\\
\tableline

      &    &              &[pc]      &[$M_\odot$]           
&[km/s]            &[Myr] &[stars/pc$^3$] 
&[$M_\odot$] &[$M_\odot$] &[pc] 
&[Myr] &[Myr]\\

\tableline

A   &575   &4642   &0.450    &0.38 &6.8  &0.23 &4.8 
&3746 &7492 &0.450 &0.045 &0.60\\

B   &1247  &4298   &0.206    &0.42 &10.8 &0.066 &5.8
&4170 &8340 &0.206 
&0.021 &0.60\\
\tableline

\end{tabular}
\end{center}
\end{minipage}
}
\caption{Initial cluster models based on the expanding models allowed
by Kroupa (2000a). Note: $N=N_{\rm sing}+2\,N_{\rm
bin}<N_{\rm in}=10^4$ because pre-main-sequence eigenevolution leads
to some close binaries merging; the average stellar mass,
$<\!\!m\!\!>$, differs because different random number seeds are used
and because some stars merge; $\sigma_{\rm 3D}$: 3D velocity
dispersion of systems with $R\le3.2$~pc; $t_{\rm
cross}=2\,R_{0.5}/\sigma_{1D}$: nominal crossing time; $\rho_{\rm C}$:
central number density; $M_{\rm st}, M_{\rm g}$: mass in stars and
gas, respectively; $\tau_{\rm M}$: time-scale for gas expulsion
(eq.~\ref{eq:Mt}); $t_{\rm D}$: onset of gas-expulsion; the stellar
and gas distributions have half-mass radii $R_{0.5}, R_{0.5, {\rm g}}
(R_{\rm pl,g}(0)=0.766\,R_{0.5,{\rm g}})$, respectively.}
\label{tab:mods}
\end{table}

\vfill

\begin{table}
{\small
\begin{minipage}[t]{9cm}
\vskip 5mm

\begin{center}
\begin{tabular}{cccccc}

{\bf Model~A:} $R\le15$~pc\\
&BD &M &K &IM &O\\
\tableline

$n$  &371    &966    &230    &177    &0\\
$f$  &0.31   &0.56   &0.59   &0.66   &0\\
($n$ &216    &828    &230    &168    &12)\\
($f$ &0.84   &0.91   &0.89   &0.88   &0.78)\\

\tableline

\\

{\bf Model~B:} $R\le15$~pc\\
&BD &M &K &IM &O\\
\tableline

$n$  &771    &1486    &378    &241    &0\\
$f$  &0.17   &0.38    &0.49   &0.44   &0\\
($n$ &415    &1417    &378    &270    &30)\\
($f$ &0.63   &0.80    &0.82   &0.78   &0.77)\\  
\tableline

\end{tabular}
\end{center}
\end{minipage}
}
\caption{The number of systems and their binary proportion within
$R\le15$~pc (the approximate tidal radius) at $t=100$~Myr
(Fig.~\ref{fig:mfn100}).  The systems are subdivided into five
mass-ranges with primaries having $0.01-0.08\,M_\odot$ (BD),
$0.08-0.5\,M_\odot$ (M), $0.5-1.0\,M_\odot$ (K), $1.0-8.0\,M_\odot$
(IM), and $>8.0\,M_\odot$ (O). The brackets contain scaled numbers of
systems at $t=0$ (from the thin solid histogram in
Fig.~\ref{fig:mfn09}). The scaling is relative to the number of ``K''
dwarfs. The total number of systems at $t=0$ and with $R\le2.5$~pc is
5165 (model~A) and 5545 (model~B).  }
\label{tab:100Myr_15pc}
\end{table}

\vfill

\begin{table}
{\small
\begin{minipage}[t]{9cm}
\vskip 5mm

\begin{center}
\begin{tabular}{cccccc}

{\bf Model~A:} $R\le2$~pc\\
&BD &M &K &IM &O\\
\tableline

$n$  &26    &96    &27    &41    &0\\
$f$  &0.19  &0.52  &0.67  &0.66  &0\\
($n$ &25    &97    &27    &20    &1)\\
($f$ &0.84  &0.91 &0.89   &0.88  &0.78)\\

\tableline

\\

{\bf Model~B:} $R\le2$~pc\\
&BD &M &K &IM &O\\
\tableline

$n$  &81    &151    &44    &55    &0\\
$f$  &0.12  &0.44   &0.57  &0.47  &0\\
($n$ &48    &165    &44    &31    &3)\\
($f$ &0.63  &0.80   &0.82  &0.78  &0.77)\\  
\tableline

\end{tabular}
\end{center}
\end{minipage}
}
\caption{The number of systems and their binary proportion within
$R\le2$~pc at $t=100$~Myr (Fig.~\ref{fig:mfn100c}).  Otherwise as
Table~\ref{tab:100Myr_15pc}. }
\label{tab:100Myr_2pc}
\end{table}

\clearpage
\newpage 

\clearpage
\newpage 

\clearpage
\newpage 

\clearpage
\newpage 

\clearpage
\newpage 

\clearpage
\newpage 

\clearpage
\newpage 

\clearpage
\newpage 

\clearpage
\newpage 

\clearpage
\newpage 

\clearpage
\newpage 

\clearpage
\newpage 

\clearpage
\newpage 

\clearpage
\newpage 

\begin{references}

\reference{} Aarseth S.J., 1999, PASP, 111, 1333

\reference{} Aarseth S.J., 2000, Gravitational N-Body Simulations, in prep.

\reference{} Aarseth S.J., H\'enon M., Wielen R., 1974, A\&A, 37, 183

\reference{} Aarseth S.J., Lin D.N.C., Papaloizou J.C.B., 1988, ApJ,
       324, 288 

\reference{} Adams F.C., 2000, ApJ, 542, 964

\reference{} Ahmad A., Cohen L., 1973, J.Comput.Ph., 12, 389

\reference{} Bate M.R., Clarke C.J., McCaughrean M.J., 1998, MNRAS,
	297, 1163

\reference{} Binney J., Tremaine S., 1987, Galactic Dynamics,
        Princeton University Press, Princeton

\reference{} Binney J., Merrifield M., 1998, Galactic Astronomy,
        Princeton University Press, Princeton

\reference{} Bonnell I.A., Bate M.R., Zinnecker H., 1998, MNRAS, 298, 93

\reference{} Bouvier J., Rigaut F., Nadeau D., 1997, A\&A, 323, 139

\reference{} Churchwell E.D., 1997, ApJ, 479, L59

\reference{} Churchwell E.D.,  1999, in The Origin of Stars and Planetary
         Systems, ed. C.J. Lada, N.D. Kylafis, NATO Science Series
         C, Vol 540, (Dordrecht: Kluwer), p.515

\reference{} Clarke C.J., Bonnell I.A., Hillenbrand L.A., 2000, in
   Protostars and Planets IV , eds. V.G. Mannings, A.P. Boss,
   S.S. Russell (Tucson: Univ. Arizona Press), p.151

\reference{} Duquennoy A., Mayor M., 1991, A\&A, 248, 485

\reference{} Eggen O.J., 1998, AJ, 116, 1810

\reference{} Elmegreen B.G., 1983, MNRAS, 203, 1011

\reference{} Elmegreen B.G., Efremov Yu.N., 1997, ApJ, 480, 235

\reference{} Elmegreen B.G., Efremov Yu.N., Pudritz R.E., Zinnecker H., 
   2000, in Protostars and Planets IV , eds. V.G. Mannings,
   A.P. Boss, S.S. Russell
   (Tucson: Univ. Arizona Press), p.179

\reference{} Elson R.A.W., Fall S.M., Freeman K.C., 1987, ApJ, 323, 54

\reference{} de La Fuente Marcos R., 1997, A\&A, 322, 764

\reference{} Garay G., Lizano S., 1999, PASP, 111, 1049

\reference{} Geyer M.P., Burkert A., 2000, MNRAS, submitted
	(astro-ph/0007413)

\reference{} Giersz M., Spurzem R., 2000, MNRAS, 317, 581

\reference{} Gilmore G., Perryman M., Lindegren L., et al., 1998, in
 	Astronomical Interferometry, ed. R.D. Reasenberg, 
	Proc. SPIE Vol. 3350, p. 541 

\reference{} Goodwin S.P., 1997a, MNRAS, 284, 785

\reference{} Goodwin S.P., 1997b, MNRAS, 286, 669

\reference{} Hambly N.C., Hodgkin S.T., Cossburn M.R., Jameson R.F.,
         1999, MNRAS, 303, 835

\reference{} Heggie D.C., Aarseth S.J., 1992, MNRAS, 257, 513

\reference{} Henney W.J., O'Dell C.R., 1999, AJ, 118, 235

\reference{} Hillenbrand L.A., 1997, AJ, 113, 1733

\reference{} Hillenbrand L.A., Hartmann L.W., 1998, ApJ, 492, 540

\reference{} Hills J.G., 1980, ApJ, 225, 986

\reference{} Hurley J.R., 2000, PhD Thesis, Trinity College, 
	University of Cambridge

\reference{} Hurley J.R., Pols O.R., Tout C.A., 2000, MNRAS, 315, 543

\reference{} Hurley J.R., Tout C.A., Aarseth S.J., Pols O.R., 2000b,
	MNRAS, in prep.
 
\reference{} Jones B.F., Walker M.F., 1988, AJ, 95, 1755

\reference{} Klessen R.S., Burkert A., Bate M.R., 1998, ApJ, 501, L205

\reference{} K\"ahler H., 1999, A\&A, 346, 67

\reference{} K\"ohler R., Leinert Ch., 1998, A\&A, 331, 977

\reference{} Kroupa P., 1995a, MNRAS, 277, 1491

\reference{} Kroupa P., 1995b, MNRAS, 277, 1507

\reference{} Kroupa P., 1995c, MNRAS, 277, 1522

\reference{} Kroupa P., 2000a, NewA, 4, 615

\reference{} Kroupa P., 2000b, in ASP Conf. Ser. Vol. 211, Massive Stellar
         Clusters, ed. A. Lancon, C. Boily 
         (San Francisco: ASP), p.233 (astro-ph/0001202)

\reference{} Kroupa P., 2000c, MNRAS, in press

\reference{} Kroupa P., Petr M.G., McCaughrean M.J., 1999, NewA, 4, 495 (KPM)

\reference{} Lada E.A., 1999, in The Origin of Stars and Planetary
         Systems, ed. C.J. Lada, N.D. Kylafis, NATO Science Series
         C, Vol 540, (Dordrecht: Kluwer), p.441

\reference{} Lada C.J., Margulis M., Dearborn D., 1984, ApJ, 285, 141

\reference{} Lindegren L., Perryman M.A.C., 1996, A\&AS, 116, 579

\reference{} Luhman K.L., Rieke G.H., 1999, ApJ, 525, 440

\reference{} Martin E.L., Brandner W., Bouvier J., et al., 2000, ApJ,
        543, 299

\reference{} Mason B.D., Gies D.R., Hartkopf W.I., Bagnuolo W.G., et al.,
          1998, AJ, 115, 821

\reference{} Mathieu R.D., 1983, ApJ, 267, L97

\reference{} Mathieu R.D., 1994, ARA\&A, 32, 465

\reference{} McCaughrean M.J., Stauffer J.R., 1994, AJ, 108, 1382

\reference{} Megeath S.T., Herter T., Beichman C., Gautier N., et al.,
          1996, A\&A, 307, 775

\reference{} Mermilliod J.-C., Rosvick J.M., Duquennoy A., Mayor M.,
         1992, A\&A, 265, 513

\reference{} Meusinger H., Schilbach E., Souchay J., 1996, A\&A, 312, 833

\reference{} Mikkola S., Aarseth S.J., 1993, Cel. Mech. Dyn. Astron.,
         57, 439

\reference{} Palla F., Stahler S.W., 1999, ApJ, 525, 772

\reference{} Palla F., Stahler S.W., 2000, ApJ, 540, 255

\reference{} Petr M.G., 1998, Binary Stars in the Orion Trapezium
     Cluster: A High angular Resolution Near-Infrared Imaging Study,
     PhD thesis, University of Heidelberg

\reference{} Pinfield D.J., Jameson R.F., Hodgkin S.T., 1998, MNRAS,
	299, 955

\reference{} Pinto F., 1987, PASP, 99, 1161

\reference{} Portegies Zwart S.F., McMillan S.L.W., Hut P., Makino J.,
	2000, MNRAS, submitted

\reference{} Preibisch T., Balega Yu., Hofmann K.-H., Weigelt G., 
	Zinnecker H., 1999, NewA, 4, 531

\reference{} Prosser C.F., Stauffer J.R., Hartmann L.W., Soderblom
     D.R., Jones B.F., Werner M.W., McCaughrean M.J., 1994,
     ApJ, 421, 517 

\reference{} Raboud D., Mermilliod J.-C., 1998, A\&A, 329, 101

\reference{} Richichi A., Leinert Ch., Jameson R., Zinnecker H.,
     1994, A\&A, 287, 145

\reference{} R\"oser S., 1999, Reviews in Modern Astronomy, 12, 79

\reference{} Saiyadpour A., Deiss B.M., Kegel W.H., 1997, A\&A, 322, 756

\reference{} Scally A., Clarke C., McCaughrean M.J., 1999, MNRAS, 306, 253

\reference{} Selman F., Melnick J., Bosch G., Terlevich R., 1999,
      A\&A, 347, 532

\reference{} Terlevich E., 1987, MNRAS, 224, 193

\reference{} Verschueren W., 1990, A\&A, 234, 156

\reference{} Verschueren W., David M., 1989, A\&A, 219, 105

\reference{} Whitworth A., 1979, MNRAS, 186, 59

\reference{}  Wilson T.L., Filges L., Codella C., Reich W., Reich P.,
     1997, A\&A, 327, 1177

\reference{} Zinnecker H., McCaughrean M.J., Wilking B.A. 1993, 
   in Protostars and Planets III, eds. E.H. Levy, J.I. Lunine,
   (Tucson: Univ. Arizona Press), p.429


\end{references}
\end{document}